\documentclass[a4paper,11pt]{article}
\pdfoutput=1 \usepackage{jheppub_2} 
\usepackage[T1]{fontenc} 
\usepackage[utf8]{inputenc}
\usepackage{array,longtable}
\usepackage{multirow}
\usepackage{booktabs}
\usepackage{mathrsfs}
\usepackage[euler]{textgreek}

\usepackage{amsmath}
\usepackage{amsthm}
\usepackage{amsfonts}
\usepackage{amssymb}
\usepackage{graphicx}
\graphicspath{ {Figures/} }
\usepackage[font={small}]{caption}
\usepackage{float}
\usepackage[export]{adjustbox}
\usepackage[hang, flushmargin,bottom]{footmisc} 
\usepackage{hyperref}
\hypersetup{
     colorlinks   = true,
     citecolor    = blue
}
\usepackage{placeins}
\usepackage{enumitem}
\usepackage{slashed}
\usepackage{bbm}
\usepackage{appendix}

\usepackage{titlesec}
\titleformat{\chapter}[display]
  {\normalfont\LARGE\bfseries}
  {\chaptertitlename\ \thechapter}{5pt}{\LARGE}
  \titlespacing*{\chapter}{0pt}{-20pt}{35pt}
\usepackage{bigstrut}
\usepackage{pst-node}
\usepackage{pstricks}
\usepackage{datetime}
\usepackage{color, colortbl}
\definecolor{GrayLight}{gray}{0.9}

\usepackage{physics}
\usepackage{mathtools}
\usepackage{setspace}
\usepackage{fancyhdr}
\usepackage[makeroom]{cancel}
\newcommand{\be}{\begin{equation}}
\newcommand{\ee}{\end{equation}}
\newcommand{\bes}{\begin{equation*}}
\newcommand{\ees}{\end{equation*}}

\renewcommand\thesection{\arabic{section}}
\usepackage{hyperref}
\hypersetup{%
  colorlinks = true,
  linkcolor  = blue
}
\usepackage{soul}
\usepackage{braket}

\usepackage{placeins}
\usepackage{hyperref}
\usepackage{makeidx}
\usepackage{mathrsfs}
\usepackage{dsfont}
\usepackage{fancybox}
\usepackage{amsfonts}
\usepackage{graphicx}
\usepackage{tcolorbox}
\usepackage{marvosym} 
\usepackage{tikz}
\usepackage{tabulary}
\usepackage{pgfplots}
\usepackage{subfig}
\usepackage{amsmath}
\usepackage{slashed}
\usepackage{mathtools}
\usepackage{fancyhdr} 
\usepackage{nccmath}
\usepackage{verbatim}
\usepackage{color}
\usepackage{amsmath, amsthm, amssymb, amsfonts}
\usepackage{array}
\newcolumntype{P}[1]{>{\centering\arraybackslash}p{#1}}
\usepackage{enumerate}
\usepackage{physics}
\usepackage[font={small, up}]{caption}
\usepackage{tikz}
\usetikzlibrary{"arrows", "automata", "backgrounds", "calendar", "chains", "matrix", "mindmap", "patterns", "petri", "shadows", "shapes.geometric", "shapes.misc", "spy", "trees"}
\usepackage{appendix}
\usepackage[normalem]{ulem}

\newcommand{\myComment}[1]{}

\newcommand{\beq}{\begin{equation}}
\newcommand{\eeq}{\end{equation}}

\newcommand{\SU}{\,{\rm SU}}

\newcommand{\U}{\,{\rm U}}

\title{ \begin{center} Enhancing $B_s \to e^+ e^-$ to an Observable Level \\ in the Two-Higgs-Doublet Model\end{center}}

\author[a]{Matthew Black,}
\author[b]{Alexis D. Plascencia,}
\author[a,c]{Gilberto Tetlalmatzi-Xolocotzi}
\affiliation[a]{Theoretische Physik 1, Center for Particle Physics Siegen (CPPS), Universit\"{a}t Siegen, Walter-Flex-Str. 3, 57068 Siegen, Germany}
\affiliation[b]{INFN, Laboratori Nazionali di Frascati, C.P. 13, 100044 Frascati, Italy}
\affiliation[c]{Université Paris-Saclay, CNRS/IN2P3, IJCLab, 91405 Orsay, France}%

\emailAdd{matthew.black@uni-siegen.de}
\emailAdd{alexis.plascencia@lnf.infn.it}
\emailAdd{gtx@physik.uni-siegen.de}

\abstract{
As a result of the helicity suppression effect, within the Standard Model the rare decay channel $B_s \to e^+ e^-$ has a decay probability which is five orders of magnitude below current experimental limits.
Thus, any observation of this channel within the current or forthcoming experiments will give unambiguous evidence of Physics Beyond the Standard Model. 
In this work, we present for the first time a New Physics scenario in which the branching fraction  $\bar{\mathcal{B}}r(B_s \to e^+ e^-)$ is enhanced up to values which saturate the current experimental bounds.
More concretely, we study the general Two-Higgs-Doublet Model (2HDM) with a pseudoscalar coupling to electrons unsuppressed by the electron mass. Furthermore, we demonstrate how this scenario can arise from a UV-complete theory of quark-lepton unification that can live at a low scale. 
This latter step allows us to establish correlations between  $B_s \to e^+ e^-$ and the lepton-flavour-violating decays $\tau^{-}\to \mu^{-}e^+ e^-$ and $\tau\to \mu \gamma$.}

\begin{document} 

\vspace*{-0.7truecm}
\begin{flushright}

SI-HEP-2022-23

P3H-22-092
\end{flushright}

\maketitle


\flushbottom

\newpage 

\section{Introduction}

The rare decays $B_s \to \ell^+ \ell^-$ for $\ell=e,\mu,\tau$ are characterized by interesting 
properties which make them quite special and suitable to test the Standard Model (SM) and to search
for New Physics (NP). For instance within the SM these transitions are only possible as loop-induced
processes. 
Moreover, they are extremely clean since only leptons are present in the final state and
all of the non-perturbative hadronic effects are contained in the decay constant of the initial $B_s$ meson. As a matter of fact the $B$ meson 
decay constants are currently known with a precision of less than $1\%$ \cite{Aoki:2021kgd, Bazavov:2017lyh, 
ETM:2016nbo, Dowdall:2013tga, Hughes:2017spc}.  

One of the particular features of the $B_s$ meson rare processes is that their decay probability in the SM is proportional to the mass of the final state lepton; this is the so-called helicity suppression effect. For muons in the final state, this leads to a SM branching fraction of $\bar{\mathcal{B}}r(B_s\rightarrow \mu^+\mu^-)=(3.55\pm 0.10)\times 10^{-9}$ which in spite of being already quite small has being measured by different experimental collaborations leading to a combined result which is in good agreement with the theoretical determination  \cite{LHCb:2021awg,LHCb:2021vsc,ATLAS:2018cur, CMS:2022dbz}.

Due to the tiny mass of the electrons, for the channel  $B_s \to e^+ e^-$  the helicity suppression is maximal. Indeed, in the SM we have $\bar{\cal B}r(B_s \to e^+ e^-)=(8.30\pm 0.36)\times 10^{-14}$ which is about four orders of magnitude below the corresponding value for $B_s \to \mu^+ \mu^-$.  Consequently, for a long time the experimental search for this channel was not pursued. As a matter of fact, until 2020 the only experimental result available was the upper bound reported by the CDF collaboration \cite{CDF:2009ssr}, which was 
then updated by the LHCb experiment \cite{LHCb:2020pcv} with the result

\beq
\bar{\cal B}r(B_s \to e^+ e^-)<9.4\times 10^{-9}.
\eeq

Due to the large difference between the most recent experimental bounds on $\bar{\mathcal{B}}r(B_s \to e^+ e^-)$ and the corresponding SM prediction we can conclude that any observation
of this channel in the foreseeable future can only be a manifestation  of physics beyond the SM. 
Following a model-independent approach, in \cite{Fleischer:2017ltw} it was shown how the presence of NP pseudoscalar interactions could enhance the SM decay probability up to values which can potentially saturate 
the known experimental bounds.  One of the main requirements to achieve this effect is that the NP
couplings should not be proportional to the mass of the electron $m_e$. 
This then excludes models where the coupling between the NP pseudoscalars and the final state electrons  is determined at leading order by the mass $m_e$. 

In this work, we present a minimal extension of the SM based on the type-III Two-Higgs-Doublet Model (2HDM), in which a second Higgs is introduced with the same quantum numbers as the SM Higgs and both scalar doublets are coupled to quarks and leptons. We show how this scenario gives enough freedom to obtain couplings between the electrons and the relevant scalar and pseudoscalar particles that allow us to  achieve large enhancements on $\bar{\mathcal{B}}r(B_s \to e^+ e^-)$ while obeying all the relevant phenomenological constraints. The literature on 2HDM is vast, for reviews on the topic we refer the reader to Refs.~\cite{Gunion:1989we,Branco:2011iw}.  

Furthermore, we study how the type-III 2HDM scenario with the properties outlined above can arise from a UV theory of quark-lepton unification.  J. Pati and A. Salam~\cite{Pati:1974yy} postulated the idea of matter unification in which the SM quarks and leptons belong to the same multiplet and this approach remains as one of the best-motivated frameworks for physics beyond the SM. However, since the top quark Yukawa coupling is predicted to be the same as the Dirac neutrino coupling, then the seesaw mechanism~~\cite{Minkowski:1977sc,Yanagida:1979as,GellMann:1980vs,Mohapatra:1979ia} requires the energy scale associated with the theory to be very high $\sim 10^{14}$ GeV, making it hard to be phenomenologically tested. Consequently, here we consider the theory proposed in Ref.~\cite{Perez:2013osa}, which can be regarded as a low energy limit of the original Pati-Salam scenario, in which neutrinos acquire their mass through the inverse seesaw mechanism~\cite{Mohapatra:1986aw,Mohapatra:1986bd} and the theory can be realized at a low energy scale.

This paper is structured as follows. In Section~\ref{sec:Bll}, we overview the experimental and theoretical status regarding the $B_s$ meson rare decays. In Section~\ref{sec:2HDM}, we discuss the general 2HDM and study the Wilson operators generated in this model. In Section~\ref{sec:results}, we present the corresponding predictions for $\bar{\mathcal{B}}r(B_s \to e^+ e^-)$ and study the phenomenological constraints on the parameter space considering different observables. In Section~\ref{sec:2HDMQLU}, we present the theoretical motivation from a theory of quark-lepton unification. Finally, our conclusions are presented in Section~\ref{sec:summary}.

\section{$B_s$ Meson Rare Decays}
\label{sec:Bll}

In order to address the decays $B_s\rightarrow \ell^+\ell^-$ we consider the following effective Hamiltonian

\beq
\mathcal{H}_{\rm eff} = -\frac{G_F V_{tb}V_{ts}^* \alpha}{\sqrt{2}\pi} \left[
C^{\ell \ell}_{10}O^{\ell \ell}_{10} 
+ C_S^{\ell\ell} O^{\ell\ell}_S + C_P^{\ell\ell} O^{\ell\ell}_P +  
+ C_{10'}^{\ell\ell}O^{\ell\ell}_{10'}
+ C_{S'}^{\ell\ell} O^{\ell\ell}_{S'} + C_{P'}^{\ell\ell} 
O^{\ell\ell}_{P'} \right] + {\rm h.c.} \, ,
\label{eq:EffHam}
\eeq
where
\begin{align}
    \mathcal{O}^{\ell\ell}_{10}  &= \left(\bar{s} \gamma_{\mu} P_L b\right)\left(\bar{\ell}\gamma^{\mu}\gamma_5\ell\right),
                                 & 
    \mathcal{O}^{\ell\ell}_{10'} &= \left(\bar{s} \gamma_{\mu} P_R b\right)\left(\bar{\ell}\gamma^{\mu}\gamma_5\ell\right),
    \nonumber\\
    \mathcal{O}^{\ell\ell}_S     &= m_b \left(\bar{s} P_R b\right)\left(\bar{\ell} \ell\right),
                                 & 
    \mathcal{O}^{\ell\ell}_{S'}  &= m_b \left(\bar{s} P_L b\right)\left(\bar{\ell}  \ell\right),
    \nonumber\\
    \mathcal{O}^{\ell\ell}_P     &= m_b \left(\bar{s} P_R b\right)\left(\bar{\ell} \gamma^5 \ell\right),
                                 &
    \mathcal{O}^{\ell\ell}_{P'}  &= m_b \left(\bar{s} P_L b\right)\left(\bar{\ell} \gamma^5 \ell\right),
\end{align}	
for $\ell=e,\mu,\tau$.

The description of the $B_s$ meson rare decays $B_s\rightarrow \ell^+\ell^-$ is given in terms of two
measurable quantities which offer complementary information. The first
one is the time-integrated branching fraction \cite{Fleischer:2017ltw,DeBruyn:2012wk}
\begin{eqnarray}
\bar{\cal B}r(B_s\rightarrow \ell^+ \ell^-)=\frac{1}{2}\int^{\infty}_{0}\Braket{\Gamma(B_s(t)\rightarrow \ell^+ \ell^-)}dt,	
\end{eqnarray}	
and the second one is the effective lifetime
\begin{eqnarray}
\tau_{\ell\ell}&\equiv&\frac{\int^{\infty}_{0}t\Braket{\Gamma(B_s(t)\rightarrow \ell^+ \ell^-)}dt}{\int^\infty_{0}
	\Braket{\Gamma(B_s(t)\rightarrow \ell^+ \ell^-)}dt},	
\end{eqnarray}	
which is equivalent to the observable
\begin{eqnarray}
\mathcal{A}^{\ell\ell}_{\Delta \Gamma_s}&=&\frac{1}{y_s}\frac{(1-y^2_s)\tau_{\ell\ell}-(1+y^2_s)\tau_{B_s}}{2\tau_{B_s}-(1-y^2_s)\tau_{\ell\ell}}.\label{eq:ADG}	
\end{eqnarray}	
In Eq.~(\ref{eq:ADG}), $\tau_{B_s}$ refers to the lifetime of the $B_s$ meson. 
In addition, the neutral $B_s$ mixing effects are accounted for by 
\begin{eqnarray}
y_{s}&\equiv&\frac{\Delta \Gamma_s}{2\Gamma_s},
\end{eqnarray}	
where $\Delta \Gamma_s$ is the decay width difference between the $B_s$ and $\bar{B}_s$ mesons.
Moreover, the untagged rate is defined as
\begin{eqnarray}
\Braket{\Gamma(B_s(t)\rightarrow \ell^+ \ell^-)}&\equiv&
\Gamma(B^0_s(t)\rightarrow \ell^+ \ell^-) + \Gamma(\bar{B}^0_s(t)\rightarrow \ell^+ \ell^-)\nonumber\\
&=& \Gamma(B_s\rightarrow \ell^+ \ell^-)_{\rm prompt}\times e^{-t/\tau_{B_s}} \Bigl(\cosh(y_s t/\tau_{B_s})+
\mathcal{A}^{\ell\ell}_{\Delta \Gamma_s}\sinh(y_s t/\tau_{B_s}) \Bigl),\nonumber\\
\label{eq:untagged}
\end{eqnarray}	
where
\begin{eqnarray}
\Gamma(B_s\rightarrow \ell^+ \ell^-)_{\rm prompt}&=&\frac{G^2_{F}\alpha^2}{16\pi^3}|V_{ts}V^{*}_{tb}|^2f^2_{B_s}M_{B_s}m^2_{\ell}
\sqrt{1-4\frac{m^2_{\ell}}{M^2_{B_s} }}|C^{\rm SM}_{10}|^2 \Bigl( |P_{\ell\ell}|^2 +  |S_{\ell\ell}|^2 \Bigl).\label{eq:promptDeltaGs}\nonumber\\
\end{eqnarray}	
The functions $P_{\ell\ell}$ and $S_{\ell\ell}$ are given by
\begin{eqnarray}
\label{eq:PS}	
P_{\ell\ell}&\equiv& \frac{C^{\ell \ell}_{10}-C^{\ell \ell}_{10'}}{C^{\rm SM}_{10}}+
\frac{M^2_{B_s}}{2m_{\ell}}\Bigl(\frac{m_b}{m_b+m_s}\Bigl)\Bigl[\frac{C^{\ell\ell}_P-
C^{\ell\ell}_{P'}}{C^{\rm SM}_{10}}\Bigl],\nonumber\\
S_{\ell\ell}&\equiv&\sqrt{1-4\frac{m^2_{\ell}}{M^2_{B_s}}}\frac{M^2_{B_s}}{2m_{\ell}}
\Bigl(\frac{m_b}{m_b+m_s}\Bigl)\Bigl[\frac{C^{\ell\ell}_{S}-C^{\ell\ell}_{S'}}{C^{\rm SM}_{10}}\Bigl].
\end{eqnarray}	
In the SM, $C^{\ell\ell}_P=C^{\ell\ell}_{P'}=C^{\ell\ell}_S=C^{\ell\ell}_{S'}=0$, leading to 
\begin{eqnarray}
P^{\rm SM}_{\ell\ell}=1,&\quad&S^{\rm SM}_{\ell\ell}=0,
\end{eqnarray}	
thus the  branching fraction simplifies to
\begin{eqnarray}
\bar{\mathcal{B}}r(B_s\rightarrow \ell^+ \ell^-)_{\rm SM}=\frac{1}{1-y_s}\frac{G^2_{F}\alpha^2}{16\pi^3}\tau_{B_s}|V_{ts}V^{*}_{tb}|^2f^2_{B_s}M_{B_s}m^2_{\ell}
	\sqrt{1-4\frac{m^2_{\ell}}{M^2_{B_s} }}|C^{\rm SM}_{10}|^2.\label{eq:SMBr}
\end{eqnarray}

For real Wilson coefficients, the theoretical branching fraction for the process 
$B_s\rightarrow \ell^+ \ell^-$ is
\begin{eqnarray}
\bar{\mathcal{B}}r(B_s\rightarrow \ell^+ \ell^-)&=& \bar{\mathcal{B}}r(B_s\rightarrow \ell^+ \ell^-)_{\rm SM}
\times
\Bigl[|P_{\ell\ell}|^2 + \frac{1-y_s}{1+y_s}|S_{\ell\ell}|^2\Bigl].\nonumber\\
\end{eqnarray}
An analogous expression in terms of $P_{\ell\ell}$ and $S_{\ell\ell}$ can also be written for $\tau_{\ell\ell}$. However, due to the
equivalence with $\mathcal{A}^{\ell\ell}_{\Delta \Gamma_s}$ we only provide an explicit expression for the latter:
\begin{eqnarray}
\mathcal{A}^{\ell\ell}_{\Delta \Gamma_s}&=&\frac{|P_{\ell\ell}|^2-|S_{\ell\ell}|^2}{|P_{\ell\ell}|^2+|S_{\ell\ell}|^2},	
\end{eqnarray}	
and finally, $\tau_{\ell\ell}$ can be obtained by applying Eq.~(\ref{eq:ADG}).

As can be seen in Eq.~(\ref{eq:SMBr}), in the SM, the decay probability $\bar{\cal B}r(B_s\rightarrow \ell^+ \ell^-)_{\rm SM}$ is proportional to the square of the mass of the final
state lepton $m^2_{\ell}$. Since muons and electrons are particularly light, for $\bar{\mathcal{B}}r(B_s\rightarrow \mu^+ \mu^-)_{\rm SM}$ and $\bar{\mathcal{B}}r(B_s\rightarrow e^+ e^-)_{\rm SM}$ the masses $m_{\mu}$ and $m_e$ respectively act as suppression factors. Then the SM predictions for the branching fractions for the different rare decays are:

\begin{eqnarray}
    \bar{\mathcal{B}}r(B_s\rightarrow e^+ e^-)_{\rm SM}&=& (8.30\pm0.22)\times10^{-14},\label{eq:SMBrelectrons}\\
    \bar{\mathcal{B}}r(B_s\rightarrow \mu^+ \mu^-)_{\rm SM}&=&(3.55\pm0.10)\times10^{-9},\\
    \bar{\mathcal{B}}r(B_s\rightarrow \tau^+ \tau^-)_{\rm SM}&=&(7.52\pm0.20)\times10^{-7}.\label{eq:SMBrnum}
\end{eqnarray}	

For the experimental value of  $\bar{\mathcal{B}}r(B_s\rightarrow \mu^+\mu^-)$ we update the result presented in \cite{Altmannshofer:2021qrr} 
by performing a weighted average including the measurements from LHCb, ATLAS and the latest value from CMS \cite{LHCb:2021awg,LHCb:2021vsc,ATLAS:2018cur, CMS:2022dbz}: 
\begin{eqnarray}
\label{eq:bsmumu}
\bar{\mathcal{B}}r(B_s\rightarrow \mu^+ \mu^-)_{\rm Exp}=(3.39\pm 0.29)\times 10^{-9}.
\end{eqnarray}	
In addition, LHCb has performed two pioneering measurements of the effective lifetime $\tau_{\mu\mu}$
\cite{LHCb:2017rmj, LHCb:2021vsc, LHCb:2021awg}. Combining \cite{LHCb:2021vsc} and \cite{CMS:2022dbz} we obtain
\begin{eqnarray}
\tau_{\mu\mu}&=&1.83\pm  0.21~\rm{ps}.
\end{eqnarray}
For examples of studies on $B_s\rightarrow \mu^+\mu^-$ and more generically on $b\rightarrow s \ell^+\ell^-$ processes within the context of 2HDM see for example \cite{Crivellin:2013wna, Crivellin:2019dun}. 

In the case of $B_s\rightarrow \tau^+\tau^-$ the current $95\%$ C.L. bound is available \cite{LHCb:2017myy}:
\begin{eqnarray}
\bar{\mathcal{B}}r(B_s\rightarrow \tau^+\tau^-)<6.8\times 10^{-3}.
\end{eqnarray}	
Notice that according to Eq.~(\ref{eq:SMBrnum}), in the SM the decay ratio $\bar{\mathcal{B}}r(B_s\rightarrow \tau^+\tau^-)$ has the largest value amongst all final state leptons, however the reconstruction of the $\tau$ particle is a challenging task, making the experimental extraction of the corresponding decay ratio especially difficult.

Finally, due to the tiny mass of the electron, the transition $B_s\rightarrow e^+e^-$  is rather suppressed in the SM. Indeed, as can be seen in Eq.~(\ref{eq:SMBrelectrons}), this channel has the lowest branching fraction among all 
the possible leptonic final states and lies outside the reach of current or future particle physics experiments. However,  the presence of NP scalar and  pseudoscalar mediators can drastically enhance the value of 
$\bar{\mathcal{B}}r(B_s\rightarrow e^+ e^-)$ \cite{Fleischer:2017ltw}. As shown in Eq.~(\ref{eq:PS}), this effects boils down to the presence 
of the tiny factor $m_{\ell}=m_e$ 
in the denominators of the functions $P_{\ell\ell}=P_{ee}$ and $S_{\ell\ell}=S_{ee}$  which for non-zero contributions of the differences $\Delta C^{\ell\ell}_P=C^{\ell\ell}_P-C^{\ell\ell}_{P'}$ 
and $\Delta C^{\ell\ell}_S=C^{\ell\ell}_S-C^{\ell\ell}_{S'}$ can maximally lift the helicity suppression induced in the SM. In this respect, the decay channel  $B_s\rightarrow e^+e^-$ is special since its measurement in any 
foreseeable experimental facility will be a clear and unambiguous indication of NP.

In 2009, CDF reported the first bound on 
the production rate of this particular channel at $90\%$ C.L.:
\begin{eqnarray}
\bar{\mathcal{B}}r(B_s\rightarrow e^+ e^-)_{\rm Exp, CDF}<2.8\times 10^{-7}.
\end{eqnarray}	
This bound was updated recently by the LHCb collaboration \cite{LHCb:2020pcv} leading to the following $90\,(95)\%$ C.L. 
bounds:
\begin{eqnarray}
\bar{\mathcal{B}}r(B_s\rightarrow e^+ e^-)_{\rm Exp, LHCb}<9.4\,(11.2)\times 10^{-9}.	
\end{eqnarray}	
As described previously, the potential enhancement on $\bar{\mathcal{B}}r(B_s\rightarrow e^+ e^-)$ as the result of  NP scalar and
pseudoscalar particles  was first noticed in \cite{Fleischer:2017ltw} in a model-independent fashion. To the best of our knowledge an  analysis within the context of a renormalizable NP framework has not been performed so far. In the following sections we take this next step and develop a NP scenario where this effect can arise.

In order to perform the numerical 
calculations corresponding to the $B$-physics processes,  in this work we will make use of the flavour physics package \texttt{flavio}\footnote{\href{https://flav-io.github.io/}{https://flav-io.github.io/}} \cite{Straub:2018kue}. This will also allow us to combine 
observables in frequentist likelihood fits of experimental data to determine 
constraints on the parameters of our NP model.
\texttt{flavio} describes NP contributions model-independently using Effective Field Theories (EFTs) where NP enters as additions to the Wilson coefficients of the operators of the EFT.
Of interest here is the Weak Effective Theory (WET) with five active flavours (defined at the scale $m_b$), where we can directly describe the contributions from our model in the language of the relevant Wilson coefficients as laid out below.

\section{The General 2HDM and the process $B_s\rightarrow e^+e^-$ }
\label{sec:2HDM}

In this Section, we will focus on a mechanism that lifts the helicity suppression in $\bar{\mathcal{B}}r(B_s \to e^+ e^-)$ leading to a large enhancement within a minimal extension of the SM. Namely, extending
the SM with a second Higgs doublet with the same quantum numbers as the SM one; for 
reviews on the Two-Higgs-Doublet Model (2HDM) we refer the reader to Refs.~\cite{Gunion:1989we,Branco:2011iw}. In the general 2HDM, both Higgs doublets are coupled to the 
quarks and leptons; this scenario is commonly referred to in the literature as the 
type-III 2HDM. Therefore, we can write the following Yukawa interactions,
\begin{align}
-\mathcal{L} &\supset \bar{Q}_L \left( Y^u_1 \widetilde{H}_1 +  Y^u_2 \widetilde{H}_2\right)u_R + \bar{Q}_L \left( Y^d_1 H_1 +  Y^d_2 H_2 \right) d_R \nonumber \\[1.5ex]
	& + \bar{\ell}_L \left( Y^e_1 H_1 + Y^e_2 H_2 \right) e_R + {\rm h.c.} \, ,
\end{align}
with $H_1^T=(H_1^+,(v_1 + H_1^0 + i A_1^0)/\sqrt{2})$, $\widetilde{H}_1=i\sigma_2 H_1^*$ and correspondingly for $H_2$. The vacuum expectation values (vevs) are defined by $\langle H_1^0 \rangle = v_1$ and $\langle H_2^0 \rangle = v_2$.

The scalar potential for $H_1$ and $H_2$ with quantum numbers $(\mathbf{1}, \mathbf{2}, 1/2)$ corresponds to
\begin{align}
	V(H_1, H_2) &= m_{11}^2 H_1^\dagger H_1 + m_{22}^2 H_2^\dagger H_2 - m_{12}^2 \left[\left( H_1^\dagger H_2 \right) + {\rm h.c.} \right] \nonumber \\ 
	&+ \frac{\lambda_1}{2} \left( H_1^\dagger H_1\right)^2 + \frac{\lambda_2}{2} \left( H_2^\dagger H_2 \right)^2 + \lambda_3 \left( H_1^\dagger H_1 \right) \left( H_2^\dagger H_2 \right) + \lambda_4 \left( H_1^\dagger H_2 \right) \left( H_2^\dagger H_1 \right) \nonumber \\
	&+ \left[ \frac{\lambda_5}{2} \left( H_1^\dagger H_2\right)^2 + \lambda_6 \left( H_1^\dagger H_1 \right) \left( H_1^\dagger H_2 \right) + \lambda_7 \left( H_2^\dagger H_2\right) \left( H_1^\dagger H_2 \right) + {\rm h.c.}\right].
 \label{eq:ScalarPotential}
\end{align}
The physical Higgs fields are defined by:
\begin{align}
\begin{pmatrix} H \\ h \end{pmatrix} & = \begin{pmatrix} \cos \alpha & \sin \alpha \\ -\sin \alpha & \cos \alpha \end{pmatrix}
\begin{pmatrix} H_1^0 \\ H_2^0 \end{pmatrix}, \\[1ex]
\begin{pmatrix} G \\ A \end{pmatrix} & = \begin{pmatrix} \cos \beta & \sin \beta \\ -\sin \beta & \cos \beta \end{pmatrix}
\begin{pmatrix} A_1^0 \\ A_2^0 \end{pmatrix}, \\[1ex]
\begin{pmatrix} G^\pm \\ H^\pm \end{pmatrix} & = \begin{pmatrix} \cos \beta & \sin \beta \\ -\sin \beta & \cos \beta \end{pmatrix}
\begin{pmatrix} H_1^\pm \\ H_2^\pm \end{pmatrix},
\label{eq:GaugetoPhysical}
\end{align}
where $h$ is identified as the SM-like Higgs and $H$ as an additional neutral Higgs. 
In addition, $H_i^0, H_i^{\pm}, A^0_i$ are the neutral, 
charged and CP-odd components of the Higgs doublets respectively. Finally $G, \, G^{\pm}$ are the would-be Goldstone bosons. 
The mixing angle $\beta$ is defined by the ratio of the vevs of the Higgs doublets, $\tan{\beta} = v_2/v_1$ and we use $v^2=v_1^2+v_2^2$. The couplings of $h$ are SM-like in the alignment limit $\sin{(\beta - \alpha)} \to 1$, which corresponds to $\alpha = \beta - \pi/2$ . Thus,
the interactions between the fermions and the neutral scalars can be written as
\begin{align}
-\mathcal{L} \supset & \,\, \bar{f}^i_L \left[ \frac{M^i_{\rm diag}}{v} h  + \left(- \cot \beta \frac{M^i_{\rm diag}}{v} + \frac{\Omega^i}{\sqrt{2} s_\beta} \right) \left( H\pm iA \right) \right] f^i_R + {\rm h.c.} \,,
\label{eq:Yukawas}
\end{align}
where the super index $i$ denotes the fermion flavour for $i=u,d,e$. In the equation above, the positive sign  is assigned to the field $A$ when considering couplings to the up-type 
quarks while the negative sign is considered for couplings to the down-type quarks and charged leptons. The mass matrices are given by
\begin{align}
m^i = Y_1^i \frac{v_1}{\sqrt{2}} + Y_2^i \frac{v_2}{\sqrt{2}},
\end{align}
and $M^i_{\rm diag}$ in Eq.~\eqref{eq:Yukawas} correspond to the diagonal mass matrices $M^i_{\rm diag} = V_L^{i\dagger} m^i V_R^i$ with unitary $V_L^i$ and $V_R^i$. 
Finally, the matrices $\Omega^i$ are given by 
$\Omega^i = V_L^{i\dagger} Y_1^i  V_R^i$ and are characterized
by general components.

For the charged leptons we assume the interaction with the heavy Higgs bosons to be very close to flavour diagonal:
\beq
\widetilde{Y}^\ell = -\cot\beta \frac{M^E_{\rm diag}}{v} + \frac{\Omega^\ell}{\sqrt{2}s_\beta} = \begin{pmatrix}
y_{ee} & \varepsilon & \varepsilon \\
\varepsilon & y_{\mu\mu} & \varepsilon \\
\varepsilon & \varepsilon & y_{\tau\tau}
\label{eq:Yleptons}
\end{pmatrix},
\eeq
where $\varepsilon \ll y_{jj}$. This allows us to evade the strong experimental constraints from the non-observation of 
processes which violate lepton flavour such as $\mu \to e \gamma$~\cite{MEG:2016leq}, $\mu-e$ conversion~\cite{SINDRUMII:2006dvw} and $\mu \to e e e$~\cite{SINDRUM:1987nra}. 
Since we are mostly interested in the coupling to electrons, we assume the hierarchy $y_{\mu\mu} \ll y_{ee} $ and $\varepsilon \ll y_{\tau\tau}$. As we will discuss in Section~\ref{sec:2HDMQLU}, the texture in Eq.~\eqref{eq:Yleptons} obeying the indicated 
hierarchy can be motivated by embedding 
the 2HDM in a low-energy limit of Pati-Salam unification.

Similarly for the down-type quarks, we assume the Yukawa interaction to be close to flavour-diagonal:
\beq
\widetilde{Y}^d = -\cot\beta \frac{M^D_{\rm diag}}{v} + \frac{\Omega^d}{\sqrt{2}s_\beta} = \begin{pmatrix}
y_{dd} & \varepsilon & \varepsilon \\
\varepsilon & y_{ss} & y_{bs}/2 \\
\varepsilon & y_{bs}/2 & y_{bb}
\end{pmatrix},
\label{eq:dyukawa}
\eeq
where we write $\varepsilon$ for very small numbers obeying $\varepsilon \ll y_{ij}$. Here, we have suppressed some 
off-diagonal entries in order to avoid the strong bounds coming from measurements of neutral kaon mixing. Also, we have kept     the off-diagonal entry $y_{bs}$ since this coupling mediates the process $B_s\rightarrow e^+ e^-$ at tree
level by coupling the NP scalar $H$ and pseudoscalar $A$ to the quarks $b$ and $s$. Moreover, we have assumed that $\widetilde{Y}^d_{sb}=\widetilde{Y}^d_{bs}=y_{bs}/2$, a choice that will be motivated in Section~\ref{sec:2HDMQLU}. As we shall see below, the experimental constraint 
from $B_s$ meson mixing requires this coupling to be very small.

The relevant Yukawa interactions affecting the process  $B_s\to e^+ e^-$ at tree level are
\beq
-\mathcal{L} \supset y_{ee} \, \bar{e} e H + y_{bs} \, \bar{b}  s H -  i \, y_{ee} \, \bar{e} \gamma^5 e A - i \, y_{bs} \, \bar{b} \gamma^5 s A,\label{eq:NPYukawa}
\eeq
where the assumption of no new sources of CP violation implies all couplings to be real. The expression in
Eq.~\eqref{eq:NPYukawa} will play a central role in our subsequent discussion. 

By integrating out the particles $A$ and $H$ inside Eq.~\eqref{eq:NPYukawa} we can immediately determine the Wilson 
coefficients $C^{ee}_{S^{(')}}$ and $C^{ee}_{P^{(')}}$ given in Eq.~(\ref{eq:EffHam}) in terms of the parameters of our model:
\begin{align}
	C^{ee}_S &=  \frac{y_{ee} y_{bs}}{M_H^2} \left( \frac{\sqrt{2}\pi}{m_b G_F V_{tb}V_{ts}^* \alpha} \right), \hspace{1.3cm} C^{ee}_{S'}= C^{ee}_S,\label{eq:CSModel}\\[1ex]
C^{ee}_P &= -\frac{y_{ee} y_{bs}}{M_A^2}\left( \frac{\sqrt{2}\pi}{m_b G_F V_{tb}V_{ts}^* \alpha} \right), \hspace{1cm} C^{ee}_{P'}= - C^{ee}_P,\label{eq:CPModel}
\end{align}
where $M_H$ and $M_A$ are the masses of $H$ and $A$ respectively.

Given that, according to Eq.~\eqref{eq:promptDeltaGs} and Eq.~\eqref{eq:PS}, the branching fraction depends only on the difference $\Delta C^{ee}_S=C^{ee}_S-C^{ee}_{S'}$
and that based on Eq.~\eqref{eq:CSModel} within our model $C^{ee}_S=C^{ee}_{S'}$, we can immediately see that the CP-even scalar $H$ 
does not contribute to the observable $\bar{\mathcal{B}}r(B_s \to e^+ e^-)$. However,  the CP-odd scalar (pseudoscalar) $A$  can have large effects.  A crucial  point to highlight is that in  our NP scenario, the coupling between the heavy Higgs bosons and electrons is not required to be proportional to the mass of the electron; this is of capital importance when lifting the SM helicity suppression.

The values of the masses $M_H$ and $M_A$ depend on the parameters of the 2HDM scalar 
potential $\lambda_{1\to7},m^2_{12}, \tan\beta$ and $\alpha$  (see Eq.~(\ref{eq:ScalarPotential}))  \cite{Gunion:2002zf}. 
Hence $M_H$ and $M_A$  are not independent from each other and are actually correlated. 
The parameters in the scalar potential are constrained by different theoretical conditions such as perturbativity and vacuum stability which can be combined as follows \cite{Gunion:2002zf}
\begin{align}
    0 &<  \lambda_{1,2} ~~~\;< 4, \\
    -\sqrt{\lambda_1\lambda_2} &< \lambda_3 \,~~~~~< 4, \\
    -4 &< \lambda_{4,5,6,7} < 4.
\end{align}
Therefore, to determine the allowed values for $M_A$ and $M_H$, we randomly sample through the parameter space of the scalar potential which in addition delivers the masses of the
charged scalars $M_{H^{\pm}}$. During this procedure, we use the mass of the SM Higgs boson
as a constraint, i.e. we take $M_h=M_h^{\rm SM}=125.25\pm 0.17~\rm{GeV}$ \cite{Workman:2022ynf} as well as the inequality $M_h \leq M_H $, and
fix $\sin(\beta-\alpha)=1$. 

\begin{figure}[t]
\centering
\includegraphics[width=0.5\textwidth]{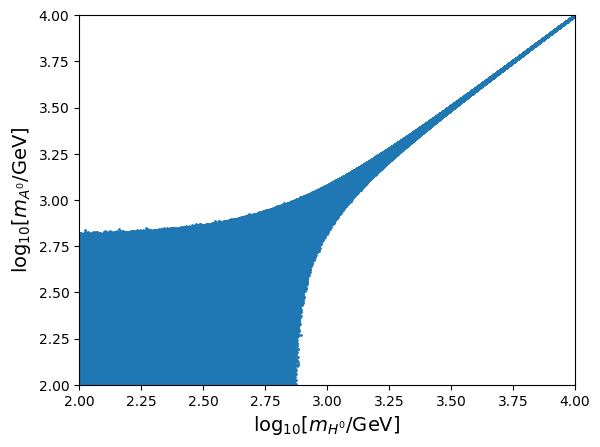}
\caption{\label{fig:MassDiffs} Correlation between the masses $M_A$ and $M_H$ from the constraints on the 2HDM scalar potential. A total of $10^9$ points fulfilling the conditions on the theory were generated, in the limit of $\sin(\beta-\alpha)=1$ and defining the perturbativity limits as $|\lambda_i|<4$. The region shaded in blue shows the masses for successful parameters of the 2HDM scalar potential.}
\end{figure}

In Fig.~\ref{fig:MassDiffs} we present our results for the allowed values of $M_A$ and $M_H$ after imposing these constraints on the parameters in the scalar potential. 
For small masses below 1 TeV, the mass splitting between $M_A$ and $M_H$ can be quite large (around 500 GeV), while for heavy masses around 10 TeV, this difference must be rather small (around 50 GeV), and hence the heavy mass regime satisfies $M_A=M_H$ to a very good approximation.
In this limit the NP Wilson coefficients given in Eq.~(\ref{eq:CSModel}) satisfy $C^{ee}_P=-C^{ee}_S$ and  $C^{ee}_{P'}=C^{ee}_{S'}$ which are two well-known relationships obtained in SMEFT. 

\newpage
\section{Enhancing $B_s \to e^+ e^-$ and Phenomenological Constraints}
\label{sec:results}
Our next task is to determine bounds for the couplings $y_{bs}$ and $y_{ee}$ to quarks and leptons respectively based on the 
phenomenological constraints available. We first focus on the bounds on the $y_{ee}$ coupling from the measurement of the cross-section for $e^-e^+\to e^-e^+$ performed by the LEP collaboration. Their reported constraints on the  four-electron axial-vector interaction~\cite{ALEPH:2006bhb} can be translated to the scalar and pseudoscalar interactions; we find that the 95\% confidence level lower bound is determined by

\beq
\frac{y_{ee}^2}{M_H^2} + \frac{y_{ee}^2}{M_A^2} < \frac{1}{( 4\,\, {\rm TeV})^2 } \,\, .
\eeq
In the case where $M_H\!=\!M_A\!$ this bound becomes 
\begin{eqnarray}
\frac{y_{ee}}{M_H} < \frac{1}{(5.7~{\rm TeV})}.
\end{eqnarray}

\begin{figure}[t]
    \centering
    \includegraphics[width=0.42\textwidth]{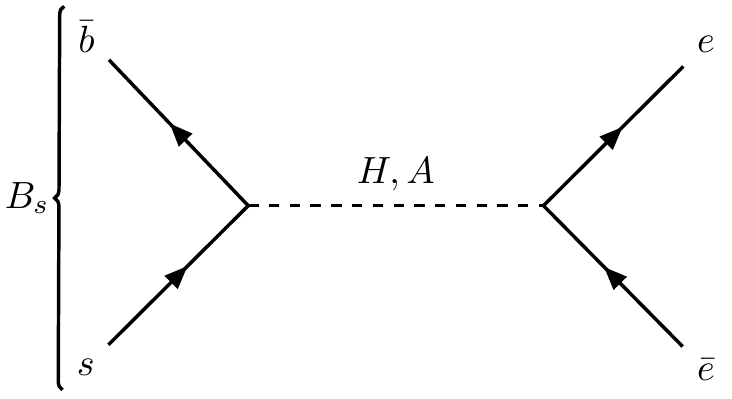} \,\,\,\,\,\,
    \includegraphics[width=0.457\textwidth]{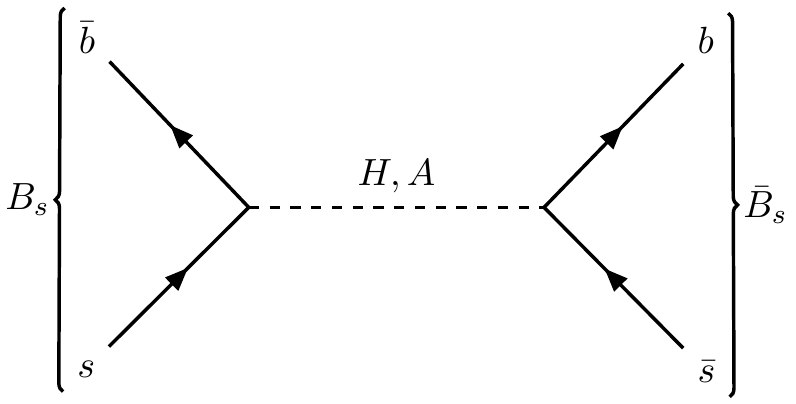}
    \caption{\label{fig:FeynmanDiag} Feynman diagrams for $B_s\to e^-e^+$ and $B_s-\bar{B}_s$ mixing induced at tree-level by the scalar $H$ and pseudoscalar $A$, respectively.}
\end{figure}

The observable $\Delta M_s$ is sensitive to the presence of NP scalar and pseudoscalar particles and thus can impose strong constraints on the coupling $y_{bs}$; the new tree-level diagrams mediated by $H$ and $A$ are shown in Fig.~\ref{fig:FeynmanDiag}. In this work we will consider the following set of $\Delta B=2$ operators which contribute to $\Delta M_s$:
\begin{align}
    \mathcal{O}^{\Delta B=2}_{V} &= \bar{s}_i\gamma^{\mu} (1-\gamma_5) b_{i}~ \bar{s}_j\gamma_{\mu}(1-\gamma_5) b_{j},
                                 &
    \mathcal{O}^{\Delta B=2}_{LL} &= \bar{s}_i (1-\gamma_5) b_{i}~\bar{s}_j (1-\gamma_5) b_{j},\nonumber\\
    \mathcal{O}^{\Delta B=2}_{RR} &= \bar{s}_i (1+\gamma_5) b_{i}~\bar{s}_j (1+\gamma_5) b_{j},
                                  &
    \mathcal{O}^{\Delta B=2}_{LR} &= \bar{s}_i (1-\gamma_5) b_{i}~\bar{s}_j (1+\gamma_5) b_{j},
\label{eq:DeltaB2}
\end{align}	
where in the SM only the coefficient of $\mathcal{O}^{\Delta B=2}_{V}$ is non-zero. In terms of the parameters of
our model in Eq.~\eqref{eq:NPYukawa}, the coefficients of the operators $\mathcal{O}^{\Delta B=2}_{LL},\mathcal{O}^{\Delta B=2}_{RR}$ and $\mathcal{O}^{\Delta B=2}_{LR}$ are respectively
\begin{align}
    C^{\Delta B=2}_{RR} &= \frac{y^2_{bs}}{4}\Bigl[\frac{1}{m^2_H} - \frac{1}{m^2_A}\Bigl], 
                        &
    C^{\Delta B=2}_{RR} &= C^{\Delta B=2}_{LL}, 
                        &
    C^{\Delta B=2}_{LR} &= \frac{y^2_{bs}}{2}\Bigl[\frac{1}{m^2_H} + \frac{1}{m^2_A}\Bigl].	\label{eq:CDeltaB2}
\end{align}	
The relevant  matrix elements of the  operators in Eq.~(\ref{eq:DeltaB2}) are given by \cite{DiLuzio:2019jyq} 
\begin{eqnarray}
\langle \mathcal{O}^{\Delta B=2}_{V} \rangle&=&\frac{8}{3}M^2_{B_{s}}f^2_{B_{s}}B_{1}(\mu_b),\nonumber
\end{eqnarray}	
\begin{eqnarray}
\langle \mathcal{O}^{\Delta B=2}_{LL} \rangle&=&M^2_{B_{s}}f^2_{B_{s}}\frac{-5 M^2_{B_s}}{3(\bar{m}_b(\mu_b)+ \bar{m}_s(\mu_b))^2}B_{2}(\mu_b),\nonumber\\
\langle \mathcal{O}^{\Delta B=2}_{LL} \rangle&=& \langle \mathcal{O}^{\Delta B=2}_{RR}\rangle,\nonumber\\
\langle \mathcal{O}^{\Delta B=2}_{LR} \rangle&=&M^2_{B_{s}}f^2_{B_{s}}\Bigl[\frac{2 M^2_{B_{s}}}{(\bar{m}_b(\mu_b)+ \bar{m}_s(\mu_b))^2} + \frac{1}{3}\Bigl] B_{4}(\mu_b). \nonumber
\end{eqnarray}	
The observable $\Delta M_s$ is calculated according to
\begin{eqnarray}
\Delta M_{s}=2|M^s_{12}|,	
\end{eqnarray}	
where
\begin{eqnarray}
M^s_{12}&=&\frac{G^2_{F}}{12\pi^2}\lambda^2_t M^2_{W}S_{0}(x_t)\hat{\eta_B}f^2_{B_s}M_{B_s}B_1+\frac{1}{2 M_{B_s}}
\Bigl[2C^{\Delta B=2}_{RR}\langle \mathcal{O}^{\Delta B=2}_{RR}\rangle +  C^{\Delta B=2}_{LR} \langle \mathcal{O}^{\Delta B=2}_{LR}\rangle \Bigl].\nonumber\\
\end{eqnarray}
To estimate $\Delta M_s$ we use \texttt{flavio}. Our inputs are the Bag parameters given in 
\cite{DiLuzio:2019jyq} and the values for $|V_{us}|,\,|V_{cb}|,\,|V_{ub}|,\gamma$ from the CKMfitter's 
Spring `21 update \cite{Charles:2004jd}. Thus, our determination in the SM is
\begin{eqnarray}
    \Delta M^{\rm SM}_{s}&=& 17.49 \pm 0.64 \, {\rm ps}^{-1}.	
    \label{eq:DMsSM}
\end{eqnarray}	

This result is in agreement with previous calculations, but its central value is noticeably lower in 
comparison; consider for instance the result reported in \cite{DiLuzio:2019jyq} which reads 
$\Delta M^{\rm SM}_{s}= 18.4^{+0.7}_{-1.2}\,$ps$^{-1}$.
This deviation is induced mainly by the update on the CKM inputs, more specifically by the  
$\sim1\sigma$ 
decrease in $|V_{cb}|$ between the results from the CKMfitter's Summer `18 report and the
one from Spring `21 \cite{Charles:2004jd}. The experimental result for $\Delta M_{s}$ is taken from \cite{Amhis:2022mac}:
\begin{eqnarray}
    \Delta M^{\rm Exp}_{s}&=& 17.765 \pm 0.006 \, {\rm ps}^{-1}.	
\end{eqnarray}	
\begin{figure}[t]
    \centering
    \includegraphics[width=0.49\textwidth]{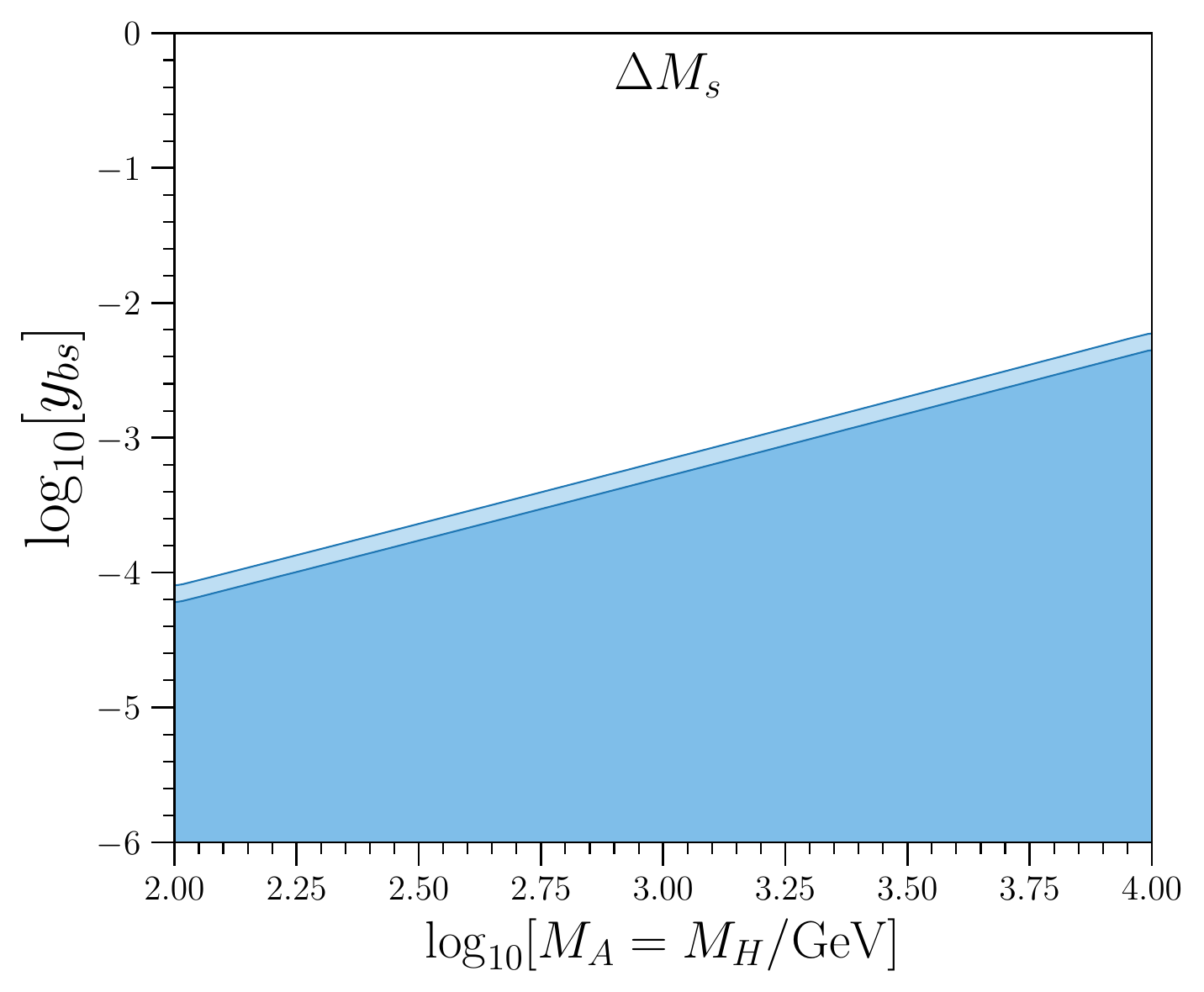}
    \includegraphics[width=0.49\textwidth]{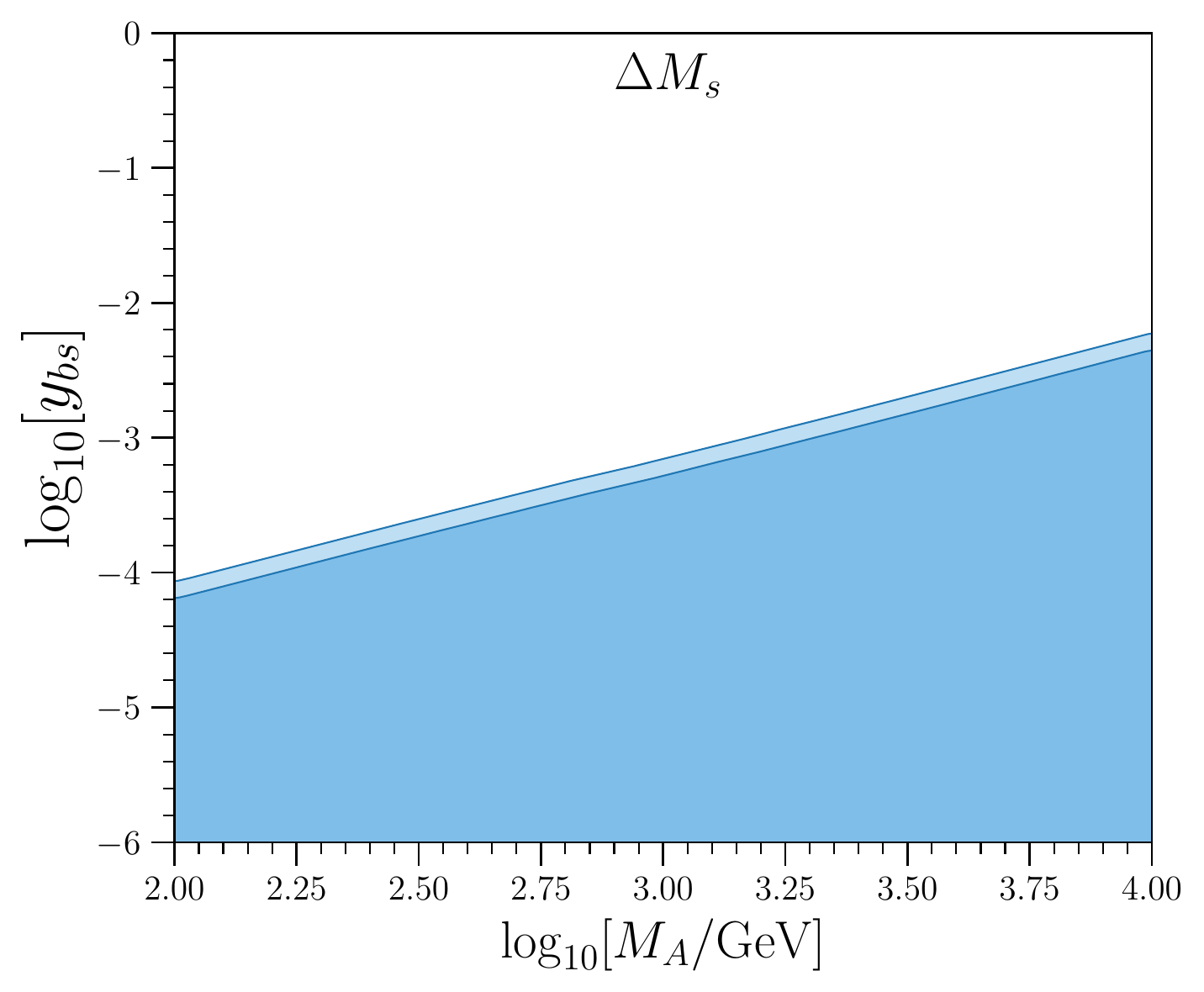}
    \caption{\label{fig:DeltaMs} Allowed parameter space of the quark coupling $y_{bs}$ and new neutral Higgs mass $M_A$, from the measurement of mass mixing in the $B_s$ system, $\Delta M_s$. 
        {\it Left panel:} In the limit of $M_A=M_H$.
        {\it Right panel:} We allow $M_A$ and $M_H$ to differ within the theoretical constraints of the model and minimize through $M_H$. 
        In both plots, the contours in dark and light blue represent the allowed space within $1\sigma$ and $2\sigma$ respectively.}
\end{figure}

In Fig.~\ref{fig:DeltaMs}, we present the constraint from the measured value of neutral $B_s$ meson mixing, $\Delta M_s$, for the allowed parameter space in the $M_H$ vs $y_{bs}$ plane, both in the limit of $M_A=M_H$ and allowing the maximum freedom between $M_A$ and $M_H$ from theory; the difference between these two scenarios is found to be minimal.
This plot shows that in order to be in agreement with the measured value of $\Delta M_s$, the coupling $y_{bs}$ has to be small; e.g. for a mass of $M_A\!=\!1\,$TeV we find that $y_{bs} \lesssim 0.001$ at $2\sigma$.

Furthermore, we can use the processes $B\rightarrow K^{(*)} e^+ e^-$ to constrain simultaneously the 
couplings $y_{bs}$ and $y_{ee}$. As a matter of fact, the NP effects in the transitions 
$B\rightarrow K^{(*)} e^+ e^-$ can be parameterized directly in terms of the Wilson coefficients 
$C^{ee}_{S^{(')}},\,C^{ee}_{P^{(')}}$ in Eq.~(\ref{eq:CPModel}) which also affect $B_s\to e^+e^-$.
The $B\to K^{(*)} e^+e^-$ observables considered in this work are listed in Table~\ref{tab:BKeeList}.
\begin{table}[th]
    \renewcommand{\arraystretch}{1.3}
    \begin{tabular}{|c|c|c|c|}
        \hline
        Observable & $q^2$ bin (GeV$^2$) & Exp. Avg. & SM Pred. \\
        \hline\hline
        $10^8\times\frac{\Delta{\cal B}}{\Delta q^2}(B^+\to K^+e^+e^-)$ & [1.0,6.0] 
        & $3.24\pm0.65$ \cite{LHCb:2014vgu,BELLE:2019xld} & $3.37\pm0.56$ \\
                                              & [0.1,4.0] 
        & $4.70\pm1.01$ \cite{BELLE:2019xld} & $3.40\pm0.58$ \\
                                              & [4.0,8.12] 
        & $2.36\pm0.79$ \cite{BELLE:2019xld} & $3.31\pm0.54$ \\[5pt]
        \hline
        $10^7\times\frac{\Delta{\cal B}}{\Delta q^2}(B^0\to K^{*0}e^+e^-)$ & [0.003,1.0] 
        & $3.09\pm0.99$ \cite{LHCb:2013pra} & $2.10\pm0.35$ \\[5pt]
        \hline
        $P_4'(B\to K^*e^+e^-)$ & [1.0,6.0] 
        & $-0.71\pm0.40$ \cite{Belle:2016fev} & $-0.34\pm0.04$ \\
                               & [14.18,19.0] 
        & $-0.15\pm0.41$ \cite{Belle:2016fev} & $-0.63\pm0.01$ \\[5pt]
        \hline
        $P_5'(B\to K^*e^+e^-)$ & [1.0,6.0] 
        & $-0.23\pm0.41$ \cite{Belle:2016fev} & $-0.42\pm0.09$ \\
                               & [14.18,19.0] 
        & $-0.86\pm0.34$ \cite{Belle:2016fev} & $-0.63\pm0.03$ \\
        \hline
    \end{tabular}
    \caption{List of $B\to Ke^+e^-$ observables used to constrain the couplings $y_{bs}$ and $y_{ee}$ and the masses $M_H$ and $M_A$. For $P_4'$ and $P_5'$, we consider the average of the $B^+$ and $B^0$ modes.}
    \label{tab:BKeeList}
\end{table}
Since the associated expressions for the observables in Table~\ref{tab:BKeeList} are quite lengthy, we 
refer the interested reader to the \texttt{flavio}'s documentation and code \cite{Straub:2018kue}.
Here we only quote explicitly the NP components of the helicity amplitudes for a  pseudoscalar $K$ or vector $K^{*}$ final state kaon which depend on the Wilson coefficients  $C^{ee}_{P^{(')}}$ and $C^{ee}_{S^{(')}}$  
\begin{align}
	h_S^{K^{*}} &= \frac{i\lambda(m_B^2,m_{K^*}^2,q^2)}{2}\Big(C^{ee}_{S}-C^{ee}_{S'}\Big)\, A_0(q^2), \label{eq:hSvt} \\
        h_P^{K^{*}} &= \frac{i\lambda(m_B^2,m_{K^*}^2,q^2)}{2}\Big(C^{ee}_{P}-C^{ee}_{P'} + \dots \Big)\, A_0(q^2),  \label{eq:hPvt}\\[1ex]
    h_S^{K} &= \frac{m_B^2-m_K^2}{2}\Big(C^{ee}_{S}+C^{ee}_{S'}\Big)\, f_0(q^2), \label{eq:hSps} \\
    h_P^{K} &= \frac{m_B^2-m_K^2}{2}\Big(C^{ee}_{P}+C^{ee}_{P'} + \dots \Big)\, f_0(q^2), \label{eq:hPps}
\end{align}
where the ellipses stand for extra contributions including the purely SM ones in $C_{10^{(')}}^{ee}$. Moreover, $\lambda(a,b,c)$ is the K\"{a}llen function, and $A_0(q^2),\,f_0(q^2)$ are each one of the $B\to K^*$ and $B\to K$ form factors respectively which depend on the invariant dilepton mass squared $q^2$ and are constructed using \cite{Bailey:2015dka,Horgan:2015vla,Bharucha:2015bzk,Gubernari:2018wyi}.
In Eqs.~\eqref{eq:hSvt} and \eqref{eq:hPvt} we can see that the NP contributions enter in terms of the differences of $\Delta C^{ee}_{S}=C^{ee}_S-C^{ee}_{S'}$ and $\Delta C^{ee}_{P}=C^{ee}_P-C^{ee}_{P'}$ as is the case for $B_s\to e^+e^-$ and therefore the $B\to K^*$ modes will only be sensitive to $M_A$. 
Conversely, from Eqs.~\eqref{eq:hSps} and \eqref{eq:hPps}, instead of the differences, the NP effects enter in terms of the sum of the relevant Wilson coefficients and so the $B\to K$ modes depend only on $M_H$.

\begin{figure}[t]
    \includegraphics[width=0.49\textwidth]{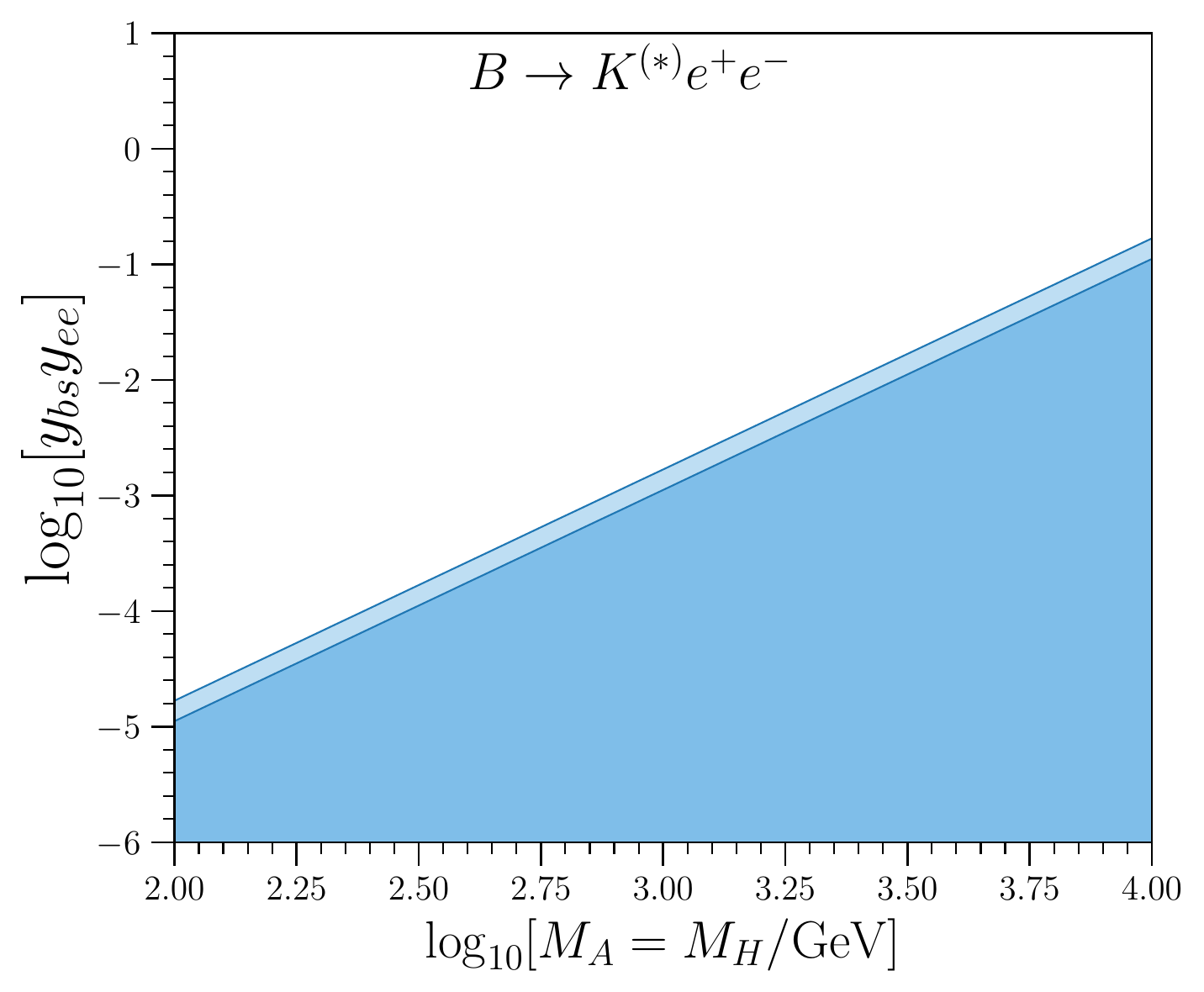}
    \includegraphics[width=0.49\textwidth]{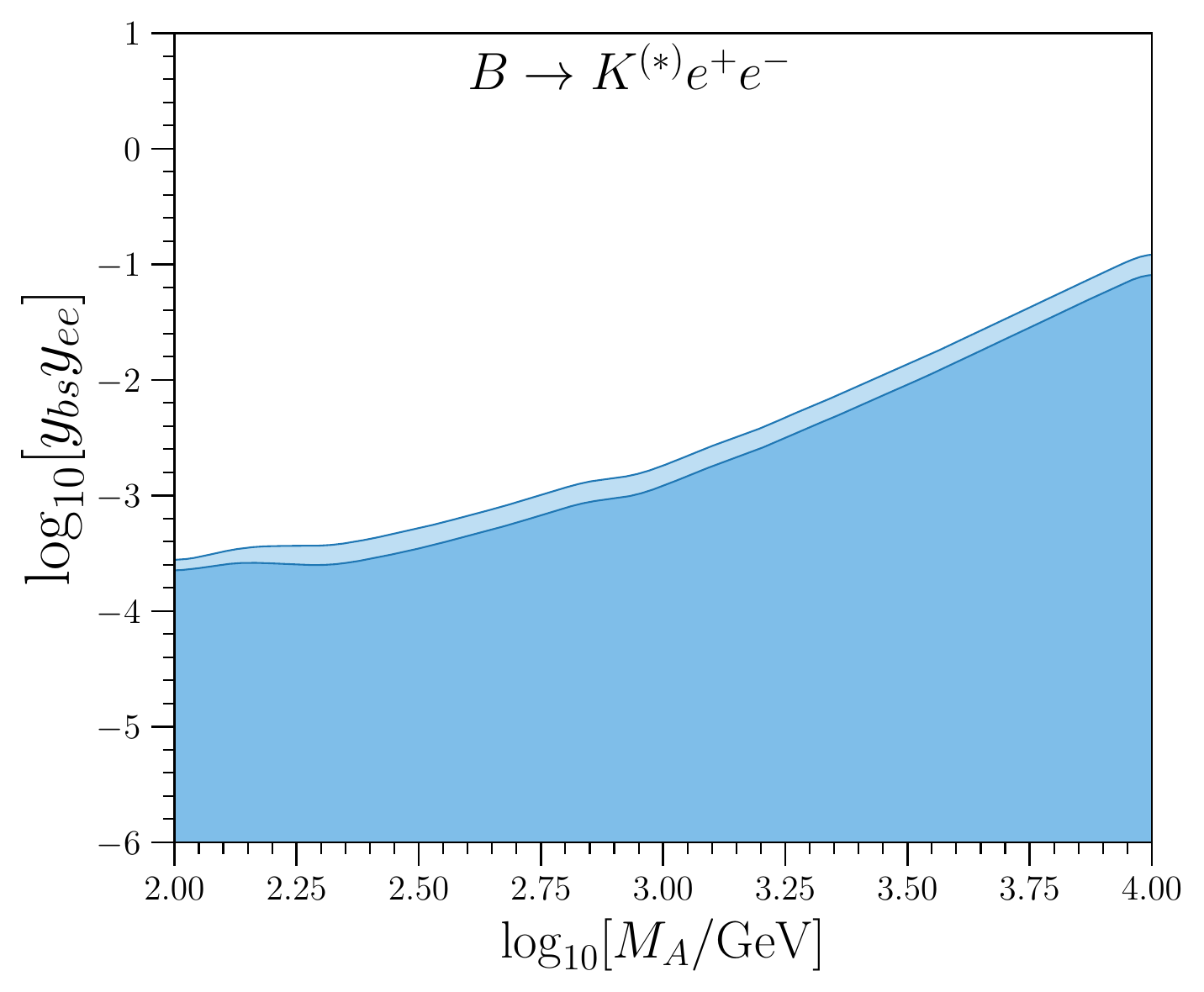}
    \caption{\label{fig:BKee} Allowed parameter space for the coupling product $y_{bs}y_{ee}$ and the new neutral Higgs mass $M_A$, from the measurements of the $B\to K^{(*)} e^+ e^-$ observables in Table~\ref{tab:BKeeList}.
        {\it Left panel:} In the limit of $M_A=M_H$.
        {\it Right panel:}  We allow $M_A$ and $M_H$ to differ within the theoretical constraints of the model and minimize through $M_H$. 
        In both plots, the contours in dark and light blue represent the possible space within $1\sigma$ and $2\sigma$ respectively.}
\end{figure}
In Fig.~\ref{fig:BKee}, we show the constraints arising from the combined fit of the $B\to K^{(*)}$ observables listed in Table~\ref{tab:BKeeList}, both in the limit of $M_A=M_H$ and allowing the maximum freedom between $M_A$ and $M_H$ from theory.
We can see that the product of the couplings $y_{bs}y_{ee}$ is expected to be small and is correlated with $M_A$ and $M_H$ similarly to the results drawn from $\Delta M_s$ for $y_{bs}$.

\begin{figure}[th]
    \centering
    \includegraphics[width=0.49\textwidth]{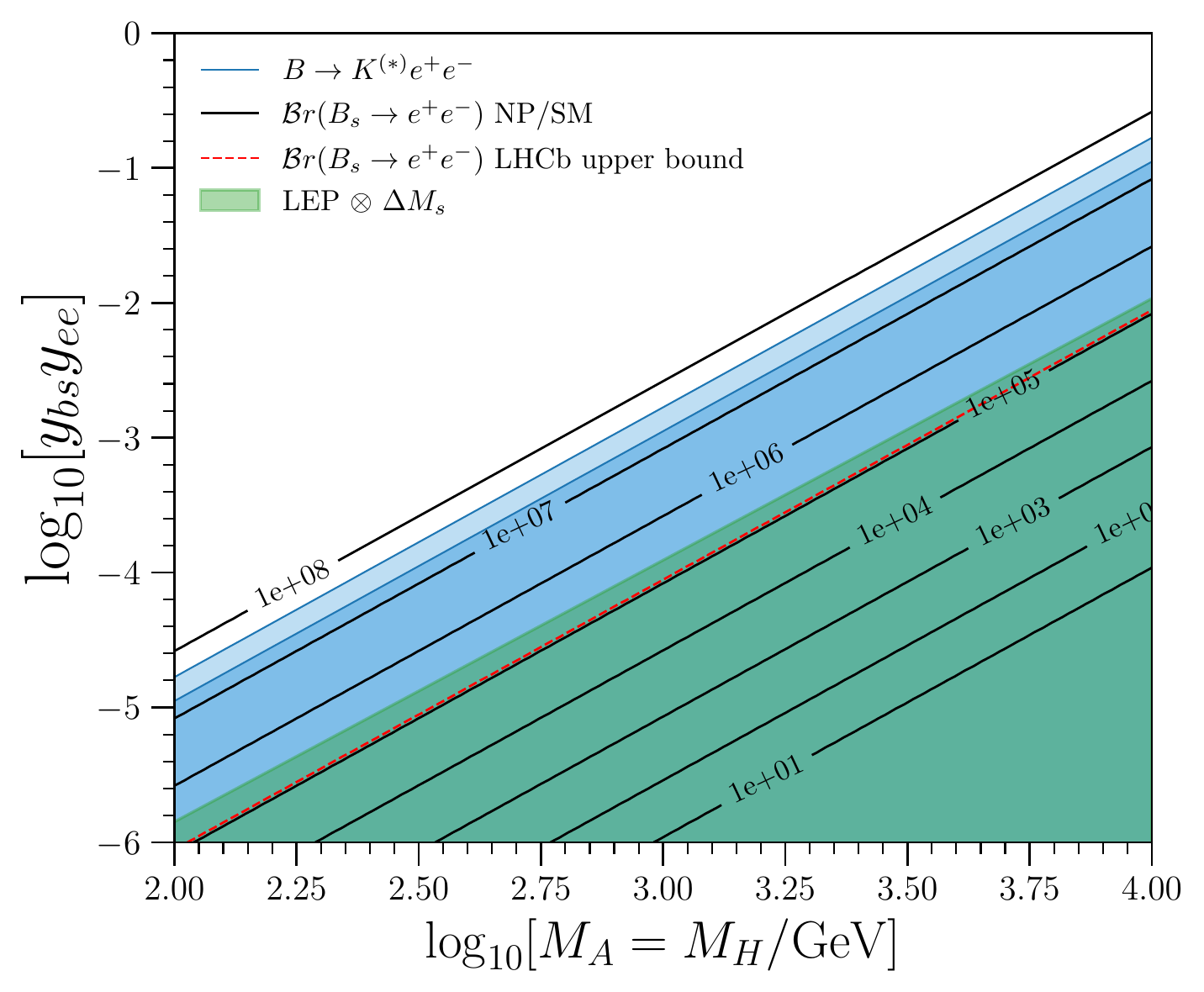}
    \includegraphics[width=0.49\textwidth]{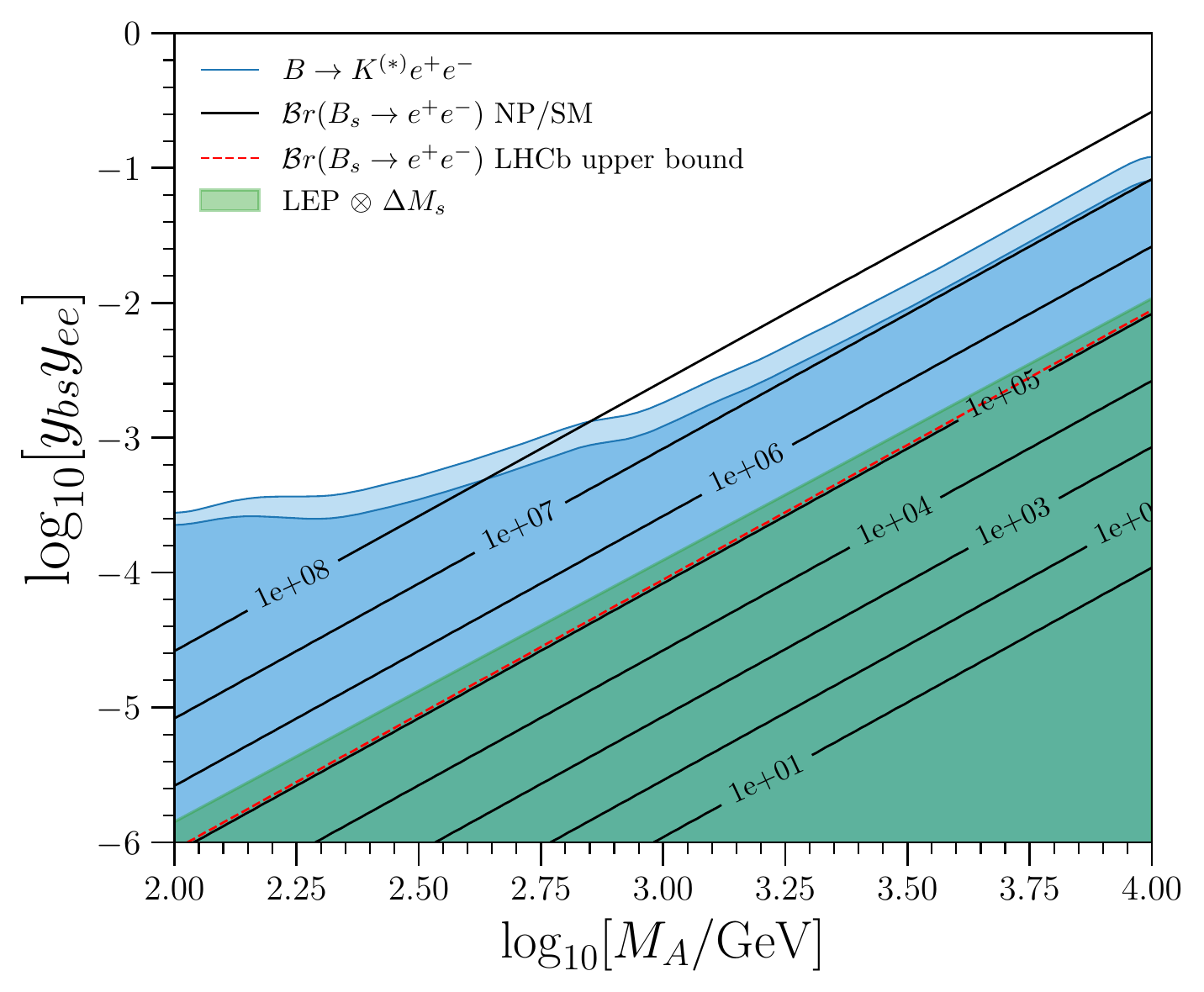}
    \caption{\label{fig:all} Allowed parameter space for the coupling product $y_{bs}y_{ee}$ and the mass of the new neutral Higgs mass $M_A$. 
        {\it Left panel:} In the limit of $M_A=M_H$.
        {\it Right panel:}  We allow $M_A$ and $M_H$ to differ within the theoretical constraints of the model and minimize through $M_H$.
        In both plots, the contours in dark and light blue represent the possible space within $1\sigma$ and $2\sigma$ respectively. 
        The black lines correspond to contours for the ratio $\bar{\cal B}r(B_s\to e^+ e^-)/\bar{\cal B}r(B_s\to e^+ e^-)_{\rm SM}$. Notice that to saturate the current experimental bound it is required an enhancement of $10^{5}$.}
\end{figure}
In Fig.~\ref{fig:all} we present the allowed parameter space in the $y_{bs}y_{ee}-M_A$ plane from all constraints considered, where once more, we take into account two cases: first the limit $M_A=M_H$ and second the situation where the maximum freedom between $M_A$ and $M_H$ is allowed from theory. 
The region shaded in green is allowed by the LEP and the $\Delta M_s$ bounds. 
The region shaded in blue is allowed by the bound from the $B\to K^{(*)}e^+e^-$ processes. 
The black lines correspond to contours for constant values of the ratio $\bar{\cal B}r(B_s\to e^+ e^-)/\bar{\cal B}r(B_s\to e^+ e^-)_{\rm SM}$ which determine the enhancement in $\bar{\cal B}r(B_s\to e^+ e^-)$ with respect to the SM prediction. 
Here we can see how an enhancement by a factor as large as $10^8$ is allowed by the collider and $B$ physics constraints. 
In fact, we can saturate the bound imposed by the LHCb analysis of $B_s \to e^+ e^-$ reported in \cite{LHCb:2020pcv} which is shown by the red dashed line and requires an enhancement by a factor of $10^5$.

Thus Fig.~\ref{fig:all} contains one of the main results of this work: within 
the context of a type-III 2HDM, an enhancement on
$\bar{\cal B}r(B_s\to e^+ e^-)$ up to values which saturate the current experimental bounds
is completely allowed and consistent with the different phenomenological constraints known
from $B$ physics and collider studies. Notice that in this Section we have focused on 
determining
the possible values that the coupling constants and masses of the type-III 2HDM affecting directly $B_s\to e^+ e^-$ can assume, although so far we have not discussed how a model with such properties could arise from a UV-complete theory. This is precisely the task we undertake in the next section. 

%
%


\section{2HDM and Quark-Lepton Unification}
\label{sec:2HDMQLU}

In this Section, we demonstrate how the coupling structure we have considered for the 2HDM can be obtained from a UV theory of quark-lepton unification that can live at a low energy scale. We focus on the NP framework proposed in Ref.~\cite{Perez:2013osa} which is based on the gauge group $$\SU(4)_C \otimes \SU(2)_L \otimes \U(1)_R.$$ Moreover, it implements the inverse seesaw mechanism in order to generate neutrino masses and can be seen as a low energy limit of the Pati-Salam theory~\cite{Pati:1974yy}. The phenomenology of the leptoquarks in this NP framework has been studied in~\cite{FileviezPerez:2021lkq,FileviezPerez:2022rbk}, while the phenomenology of its scalar sector, which corresponds to a special case of the type-III 2HDM, has been recently analyzed in~\cite{FileviezPerez:2022fni}. For further details we refer the reader to those references.

Within our framework, the SM matter fields are unified in the following representations,
\begin{eqnarray}
F_{QL} &=&
\left(
\begin{array}{cccc}
u_r \,\,& u_g \,\,& u_b  \,\,& \nu 
\\
d_r \,\,& d_g \,\,& d_b  \,\,& e
\end{array}
\right) \sim (\mathbf{4}, \mathbf{2}, 0), \\[1ex]
F_u &=&
\left(
\begin{array}{cccc}
u^c _r \,\,& u^c_g \,\,& u^c_b \,\,& \nu^c
\end{array}
\right) \sim (\mathbf{\bar{4}}, \mathbf{1}, -1/2), 
\\[1ex]
 F_d &=&
\left(
\begin{array}{cccc}
d^c_r \,\,& d^c_g \,\,& d^c_b \,\,& e^c
\end{array}
\right) \sim (\mathbf{\bar{4}}, \mathbf{1}, 1/2).
\end{eqnarray}
and hence, the leptons can be interpreted as the fourth colour of the fermions. The Yukawa interactions for the charged fermions can be written as
\begin{align}
	-\mathcal{L}_Y \supset Y_1 F_{QL} F_u H_1 + Y_2 F_{QL}  F_u \Phi + Y_3 H_1^\dagger F_{QL} F_d + Y_4 \Phi^\dagger F_{QL} F_d +  {\rm h.c.},
\end{align}
where $H_1 \sim (\mathbf{1},\mathbf{2},1/2)$ and $\Phi \sim (\mathbf{15},\mathbf{2},1/2)$ are required to generate fermion masses in a consistent manner. The $\Phi$ field contains a second Higgs doublet $H_2$ that is coupled to all the SM fermions
\beq
\Phi = 
\left(
\begin{array} {cc}
\Phi_8 & \Phi_3  \\
\Phi_4 & 0  \\
\end{array}
\right) + \sqrt{2} \, T_4 \ H_2 \sim (\mathbf{15}, \mathbf{2}, 1/2),
\eeq
where $T_4$ is one of the generators of $\SU(4)_C$ and it is normalized as
$
T_4 =
\frac{1}{2 \sqrt{6}} \rm{diag} (1,1,1,-3).
$


Since the NP framework under consideration arises from quark-lepton unification there are only four independent Yukawa couplings (instead of eight) defining the interactions between the Higgs doublets and the SM fermions:
\begin{align}
	-\mathcal{L} &= \bar{u}_R \left( Y_1^T \widetilde{H}_1 + \frac{1}{2\sqrt{3}} Y_2^T \widetilde{H}_2\right)Q_L + \bar{N}_R \left( Y_1^T \widetilde{H}_1 - \frac{\sqrt{3}}{2} Y_2^T \widetilde{H}_2 \right) \ell_L \nonumber \\
	&+ \bar{d}_R \left( Y_3^T H_1^{\dagger} + \frac{1}{2\sqrt{3}} Y_4^T H_2^{\dagger} \right) Q_L + \bar{e}_R \left( Y_3^T H_1^{\dagger} -\frac{\sqrt{3}}{2} Y_4^T H_2^{\dagger}\right) \ell_L + {\rm h.c.} \,,
\end{align}
and the vevs are defined by $\langle H_1^0 \rangle = v_1/\sqrt{2}$ and $\langle H_2^0 \rangle = v_2/\sqrt{2}$.

As it was shown in Ref.~\cite{FileviezPerez:2022fni}, the interactions between the physical Higgs bosons and the SM down-type quarks and charged leptons are given respectively by
\begin{align}
\tilde{Y}^\ell & = \left( \tan{\beta} - 3\cot{\beta}  \right) \frac{M^E_{\rm diag}}{4v} + 3\left( \tan{\beta} + \cot{\beta}\right)\frac{V_c^{T} M^D_{\rm diag} V}{4 v}, \label{eq:ceeH} \\[1.5ex]
\tilde{Y}^d & = \left( 3\tan{\beta} - \cot{\beta}  \right) \frac{M^D_{\rm diag}}{4v} + \left( \tan{\beta} + \cot{\beta}\right)\frac{V_c^{*} M^E_{\rm diag} V^{\dagger} }{4 v}, \label{eq:cddH} 
\end{align}
where $V$ and $V_c$ are unitary matrices which contain information about the unknown mixing between quarks and leptons. In addition, $M^D_{\rm diag}$ and $M^E_{\rm diag}$ are the diagonal mass matrices for down-type 
quarks and charged leptons. From Eqs.~(\ref{eq:ceeH}) and (\ref{eq:cddH}) above we can see that the theory predicts a correlation 
between the couplings to quarks and leptons. As it was demonstrated in Ref.~\cite{FileviezPerez:2021arx}, in the regimes with $\tan\beta\gg 1$ or $\tan\beta\ll 1$ the theory predicts unique relations among the decay widths of heavy Higgs bosons that can be probed at the LHC. Consequently, we focus on these two limits.

If we assume the complex phases to vanish, the $3\times 3$ unitary matrix $V$ can be parameterized in terms of three mixing angles, which here we denote as $\theta_{12}$, $\theta_{13}$ and $\theta_{23}$, as follows 
\beq 
V = 
\begin{pmatrix} c_{12} c_{13} &  s_{12} c_{13} & s_{13}\\[1ex]
-s_{12} c_{23} - c_{12} s_{23} s_{13} \,\,\, & c_{12} c_{23} - s_{12} s_{23} s_{13} & s_{23} c_{13} \\[1ex]
s_{12} s_{23} - c_{12} c_{23} s_{13} & -c_{12} s_{23} - s_{12} c_{23} s_{13} \,\,\, & c_{23} c_{13}
\end{pmatrix} ,
\eeq
where we have used $s_{ij}$ and $c_{ij}$ as 
short notation for $\sin \theta_{ij}$ and $\cos \theta_{ij}$ respectively. An analogous expression can then also be written for $V_c$ but with primed mixing angles $s'_{ij}$ and $c'_{ij}$.  For large $\tan\beta$ and in the limit where $s_{ij}\to 1$ and $s'_{ij}\to 1$ the interactions with the charged leptons are simplified to
\beq
\tilde{Y}^\ell = \frac{\tan \beta}{4v} \begin{pmatrix} 
m_e + 3m_b & \varepsilon & \varepsilon \\
\varepsilon &  m_\mu + 3m_s & \varepsilon \\
\varepsilon & \varepsilon &  m_\tau + 3m_d
\label{eq:Ceematrix}
\end{pmatrix} ,
\eeq
which gives us the flavour-diagonal couplings with the hierarchy $y_{ee} \gg y_{\mu\mu}, y_{\tau\tau}$. This motivates our choice for the couplings in Section~\ref{sec:2HDM}. The same conclusions hold for intermediate and small values of $\tan \beta$.

For the down-type quarks and large $\tan\beta$, we get the following interaction matrix
\beq
\tilde{Y}^d =  \frac{\tan \beta}{4v} \begin{pmatrix} 
3m_d + m_\tau & \varepsilon & \varepsilon \\
\varepsilon & 3m_s + m_\mu & \varepsilon \\
\varepsilon & \varepsilon & 3m_b + m_e 
\end{pmatrix} ,
\label{eq:Cddmatrix}
\eeq
which gives us the hierarchy $y_{dd}\simeq y_{bb} \gg y_{ss}$. The same conclusion holds for intermediate and small values of $\tan \beta$. Unfortunately, due to the freedom in the coupling to the right-handed neutrinos, the theory does not predict the coupling of the Higgs bosons to the up-type quarks.

\begin{figure}[t]
    \centering
    \includegraphics[width=0.495\textwidth]{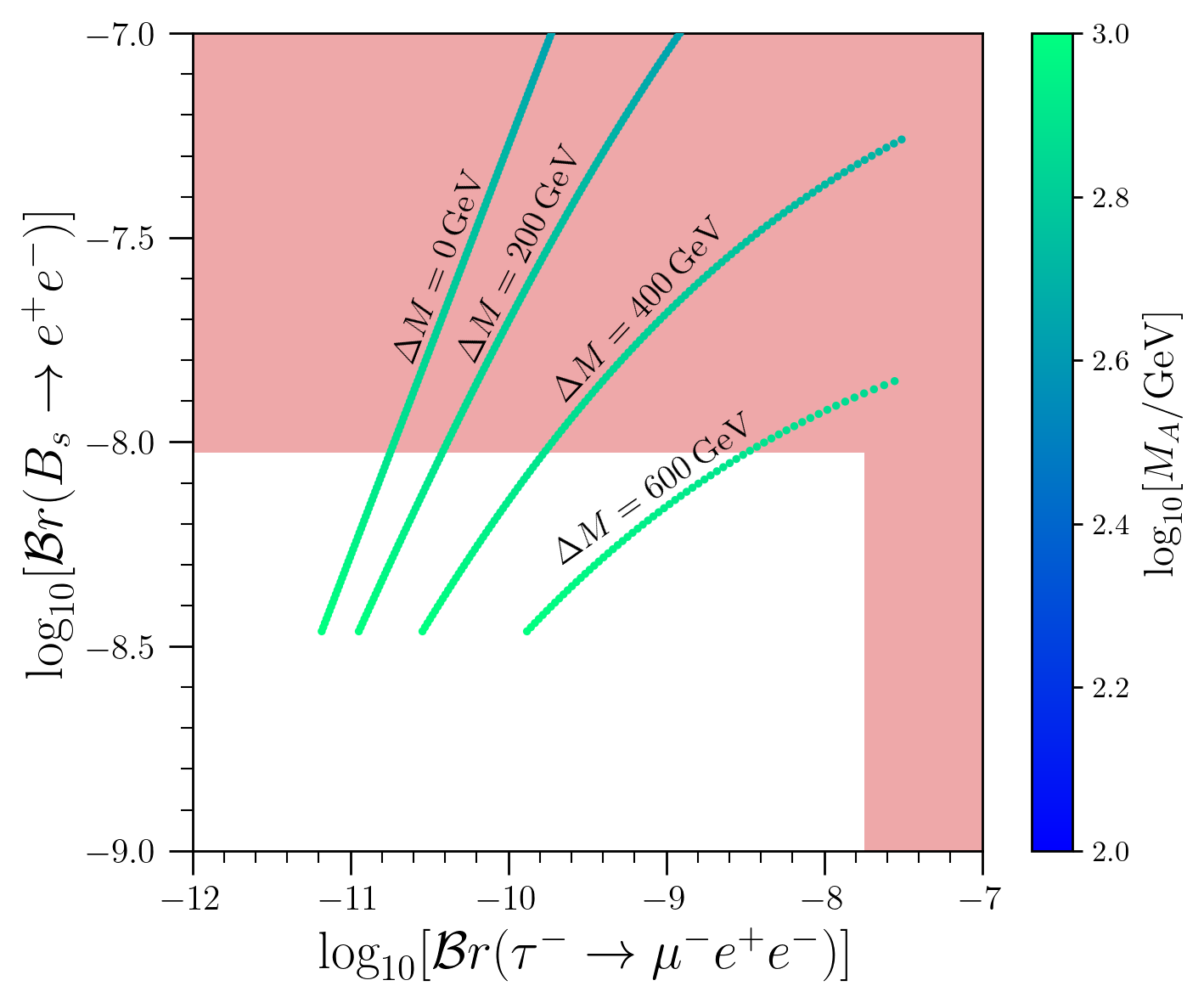}
    \includegraphics[width=0.495\textwidth]{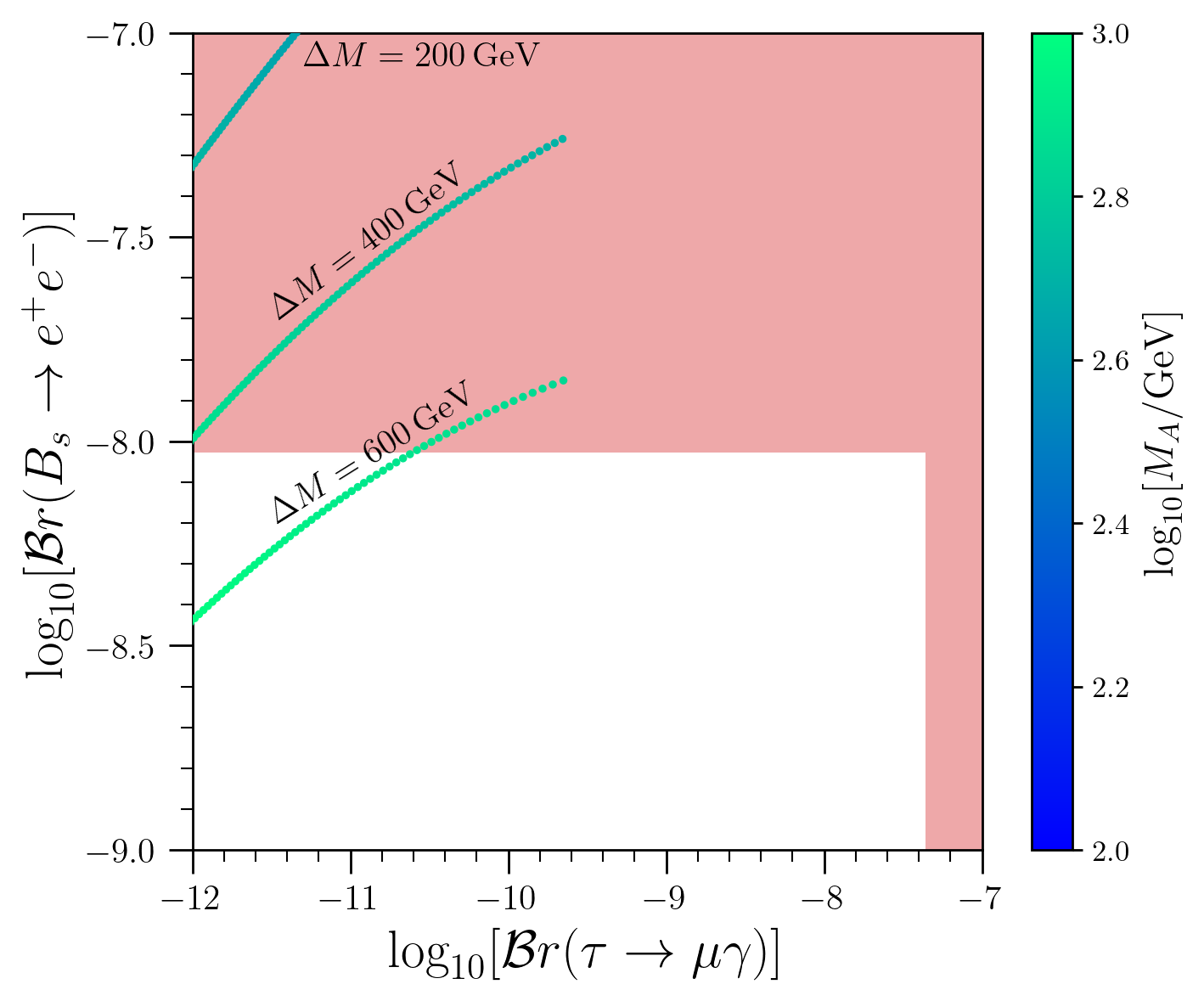}
    \caption{\label{fig:taudecays} \textit{Left panel:} Predicted correlation between $\bar{\mathcal{B}}r(B_s\to e^+e^-)$ and $\mathcal{B}r(\tau^-\to \mu^- e^+e^-)$. The regions shaded in red correspond to the experimental bounds for each decay respectively. The different bands correspond to different values for the mass splitting $\Delta M\equiv M_A-M_H$. We have fixed  $s_{23}=s'_{23}=0.98$ and $\tan\beta=10$. \textit{Right panel:} Same as the left panel for the predicted correlation between $\bar{\mathcal{B}}r(B_s\to e^+e^-)$ and $\mathcal{B}r(\tau\to \mu\gamma)$.}
\end{figure}

Since we require a non-zero $y_{bs}$ coupling, we set all $s_{ij}\!=\!s'_{ij}\!=\!1$ except for $s_{23}$ and $s'_{23}$; this choice is motivated by requiring the first-generation off-diagonal couplings to be vanishing. From quark-lepton unification, non-zero entries for $\tilde{Y}^d_{bs}$ and $\tilde{Y}^d_{sb}$  imply non-zero values for $\tilde{Y}^\ell_{\mu\tau}$ and $\tilde{Y}^\ell_{\tau\mu}$. Namely,
\begin{align}
\tilde{Y}^d_{sb} & = \frac{1}{4v} \left( \tan \beta + \cot \beta \right) \left( m_\mu s'_{23} c_{23}  - m_e s_{23} c'_{23} \right),\\[1ex]
\tilde{Y}^d_{bs} & = \frac{1}{4v} \left( \tan \beta + \cot \beta \right) \left( m_\mu s_{23} c'_{23} - m_e s'_{23} c_{23} \right),\\[1ex]
\tilde{Y}^\ell_{\mu \tau} & = \frac{3}{4v} \left( \tan \beta + \cot \beta \right) \left( m_s s'_{23} c_{23} - m_d s_{23} c'_{23} \right),\\[1ex]
\tilde{Y}^\ell_{\tau \mu} & = \frac{3}{4v} \left( \tan \beta + \cot \beta \right) \left( m_s s_{23} c'_{23} - m_d s'_{23} c_{23} \right),
\end{align}
whenever $s_{23}=s'_{23}$ then we have that $\tilde{Y}^d_{sb}=\tilde{Y}^d_{bs}=y_{bs}/2$ which motivates the choice made in Eq.~\eqref{eq:dyukawa}; we also have that $\tilde{Y}^\ell_{\mu\tau}=\tilde{Y}^\ell_{\tau\mu}=y_{\tau\mu}/2$. The $\tau\mu$ couplings will generate the following dimension-six operators
\begin{align}
\label{eq:tauH}
\mathcal{H}_{\rm eff} \supset & -\frac{y_{ee} y_{\mu\tau}}{M_H^2} (\bar{\tau}\mu) (\bar{e}e) - \frac{y_{\tau\tau} y_{\mu\tau}}{M_H^2} (\bar{\tau}\mu) (\bar{\tau}\tau) \nonumber \\[1ex]
& + 
\frac{y_{ee} y_{\mu\tau}}{M_A^2} (\bar{\tau} \gamma^5 \mu) (\bar{e} \gamma^5 e) + \frac{y_{\tau\tau} y_{\mu\tau}}{M_A^2} (\bar{\tau} \gamma^5 \mu) (\bar{\tau} \gamma^5 \tau),
\end{align}
which will induce the lepton-flavour-violating decay $\tau^\pm \to \mu^\pm e^- e^+$ at tree-level, and $\tau^\pm \to \mu^\pm \gamma$ at one-loop. The current experimental bounds on these decay channels are ${\cal B}r(\tau \to  \mu \gamma) < 4.4 \times 10^{-8}$~\cite{BaBar:2009hkt} and ${\cal B}r(\tau^-\to \mu^- \, e^+ e^-) < 1.8\times10^{-8}$~\cite{Hayasaka:2010np}. These bounds are expected to be improved by future $B$ factories~\cite{SuperB:2010cqs}.

The effective operators $\bigl\{(\bar{\tau}\mu)(\bar{e}e), (\bar{\tau}\gamma^{5}\mu)(\bar{e}\gamma^{5}e)\bigl\}$ in Eq.~\eqref{eq:tauH} can be mapped to operators which are analogous to $\mathcal{O}^{\Delta B=2}_{LL}, \mathcal{O}^{\Delta B=2}_{RR}$ and $\mathcal{O}^{\Delta B=2}_{LR}$ in Eq.~\eqref{eq:DeltaB2}, and hence, the corresponding Wilson coefficients have also analogous expressions to the coefficients shown in Eqs.~(\ref{eq:CDeltaB2}). More precisely, the new set of operators to be considered is
\begin{align}
\mathcal{O}^{ee}_{LL} &= \bar{\tau}(1-\gamma_5)\mu~\bar{e} (1-\gamma_5) e, & 
\mathcal{O}^{ee}_{RR} &= \bar{\tau}(1+\gamma_5)\mu~\bar{e} (1+\gamma_5) e, \nonumber\\
\mathcal{O}^{ee}_{LR} &= \bar{\tau}(1-\gamma_5)\mu~\bar{e} (1+\gamma_5) e, &
\mathcal{O}^{ee}_{RL} &= \bar{\tau}(1+\gamma_5)\mu~\bar{e} (1-\gamma_5) e,
\end{align}
with coefficients
\begin{align}
C^{ee}_{LL} &= C^{ee}_{RR} = \frac{y_{ee}y_{\mu\tau}}{4}\left[\frac{1}{M_H^2} - \frac{1}{M_A^2}\right],&\
C^{ee}_{LR} &= C^{ee}_{RL} = \frac{y_{ee}y_{\mu\tau}}{4}\left[\frac{1}{M_H^2} + \frac{1}{M_A^2}\right].
\end{align}
Similar expressions follow for operators constructed from the set $\bigl\{(\bar{\tau}\mu)(\tau\bar{\tau}), (\bar{\tau}\gamma^{5}\mu)(\tau\gamma^{5}\bar{\tau})\bigl\}$ with corresponding Wilson coefficients $C^{\tau\tau}_{LL},C^{\tau\tau}_{RR},
C^{\tau\tau}_{LR}$ and $C^{\tau\tau}_{RL}$. After implementing these effective operators, we proceed to compute the $\tau$-decays in \texttt{flavio}. In order to obtain a large contribution to $\tau\to \mu \gamma$, we require that $C^{\tau\tau}_{RR}$ and $C^{\tau\tau}_{LL}$ be non-zero, and hence a mass splitting between $H$ and $A$ is needed.

Our theoretical setup has effects on the  decay channel $B_s \to \mu^+\mu^-$, which depends on the Wilson coefficients $C^{\mu\mu}_S$ and $C^{\mu\mu}_P$ which are analogous to $C^{ee}_S$ and $C^{ee}_P$ for $B_s\to e^+e^-$ with the leptonic coupling $y_{ee}$ replaced by $y_{\mu\mu}$, see Eq.~(\ref{eq:CSModel}). Since we have $y_{\mu\mu} \ll y_{ee}$ this implies that the NP contribution will be much smaller for muons than for electrons. Nonetheless, we have checked that for each point in the allowed parameter space the prediction for $\bar{\cal B}r(B_s \to \mu^+\mu^-)$ is in agreement with the experimental measurement given in Eq.~\eqref{eq:bsmumu} within $2\sigma$.

Now, let us analyze the effects on $\bar{\mathcal{B}}r(B_s\rightarrow e^+ e^-)$ and the lepton-flavour-violating decays. Firstly, we provide a concrete example on how large the enhancement in ${\cal B}r(B_s \to e^+ e^-)$ can be within the theory under consideration for concrete values of the input parameters. Thus, we fix the mass of the scalar and pseudoscalar to $M_A\!=\!800$ GeV, $M_H\!=\!400$ GeV and the rest of the parameters to $\tan\beta=10$ and $s_{23}=s'_{23}=0.98$. This implies an electron coupling of $y_{ee}\simeq 0.13$ which is in agreement with the bound from LEP. For the off-diagonal quark coupling we obtain $y_{bs}\simeq 4.2 \times 10^{-4}$ which gives $\bar{\cal B}r(B_s \to e^+ e^-)\simeq 8.4\times 10^{-9}$. For the choice of mixing angles discussed above, the off-diagonal lepton coupling is $y_{\tau\mu}\simeq 1.1 \times 10^{-3}$ and predicts ${\cal B}r(\tau^-\to \mu^- \, e^+ e^-)\simeq 1.4\times 10^{-10}$ and ${\cal B}r(\tau \to \mu \gamma)\simeq 6.6\times10^{-13}$.

Finally, we can generalize the previous exercise while at the same time assessing the impact on lepton-flavour-violating decays. Then,  in the left panel of Fig.~\ref{fig:taudecays} we show the predicted correlation between the observables $\bar{\cal B}r(B_s \to e^+ e^-)$ and $\bar{\cal B}r(\tau^-\to \mu^- \, e^+ e^-)$. The different bands correspond to different values for the mass splitting $\Delta M\equiv M_A-M_H$. We fix  $s_{23}=s'_{23}=0.98$ and $\tan\beta=10$. The region shaded in red corresponds to the current experimental limits on these observables. In the right panel in Fig.~\ref{fig:taudecays} we show the correlation between $\bar{\cal B}r(B_s \to e^+ e^-)$ and ${\cal B}r(\tau \to \mu \gamma)$. From Fig.~\ref{fig:taudecays} we can see that it is possible to saturate the current experimental bounds in $\bar{\cal B}r(B_s \to e^+ e^-)$ while at the same time obeying the constraints on the lepton-flavour-violating decays. This is the second result that we want to highlight in this work. These plots provide a set of correlations between different channels which make this framework phenomenologically testable.   

\section{Summary}
\label{sec:summary}

The leptonic decay $B_s \to e^+ e^-$ is a decay channel with interesting properties and it can be used as smoking gun in the search for New Physics. For instance, it is exceptionally clean. Moreover if this process takes place as predicted by the Standard Model, due to the helicity suppression effect, its tiny decay probability places it outside the reach of current or forthcoming particle physics experiments. Therefore, any observation of this channel in the near future would represent conclusive evidence for physics beyond the Standard Model.

In this article and to the best of our knowledge, we have presented for the first time, a concrete New Physics scenario which can provide a large enhancement on the decay width for the channel $B_s \to e^+ e^-$. More specifically, by studying the general 2HDM in which both doublets are coupled to the quarks and leptons of the Standard Model, we have demonstrated that when the CP-odd scalar $A$ is mostly coupled to electrons, it can give a contribution to the transition $B_s \to e^+ e^-$ which enhances its decay probability by up to  five orders of magnitude above the Standard Model prediction, saturating the most recent experimental upper bound established by the LHCb collaboration. We have identified regions in the corresponding parameter space where this potential enhancement respects all known constraints from flavour and collider physics, including for instance neutral $B_s$ mixing as well as LEP measurements of the $e^-e^+ \to e^-e^+$ cross-section.  

Furthermore, we have shown how the required coupling structure for the 2HDM can arise from a UV theory of quark-lepton unification that can be realized at a low energy scale. This framework predicts a correlation between the decay channel 
 $B_s \to e^+e^-$ and the lepton-flavour-violating decays $\tau^-\to\mu^- e^+ e^-$ and $\tau\to \mu\gamma$. 
 We have worked out quantitatively the interplay between these channels for different values of the relevant free parameters.
 If the decay process $B_s\to e^+ e^-$ is observed in the near future, the presence of heavy (pseudo)scalars can be further confirmed by searches for a heavy resonance decaying into an electron-positron pair at the LHC. Our results show that the channel $B_s\to e^+ e^-$ is indeed a very interesting candidate to probe for New Physics effects and provide additional justification to pursue further experimental searches for it in the current and foreseeable experiments. 


\appendix

\acknowledgments
This project has received funding from the European Union’s Horizon 2020 research and innovation programme under the Marie Skłodowska-Curie grant agreement No 945422. A.D.P. is supported by the INFN “Iniziativa Specifica” Theoretical Astroparticle Physics (TAsP-LNF) and by the Frascati National Laboratories (LNF) through a Cabibbo Fellowship, call 2020.
M.B. and G.TX. are supported by Deutsche Forschungsgemeinschaft (DFG, German Research Foundation) through grant 396021762 - TRR 257 ``Particle Physics Phenomenology after the Higgs Discovery''. 
Parts of the computations carried out for this work made use of the OMNI cluster of the University of Siegen. We acknowledge useful communication with  Alexander Lenz and Matthew Kirk on the current updates of $B_s\rightarrow \mu^+ \mu-$.

\newpage
\appendix

\bibliographystyle{JHEP}
\bibliography{Bsee}{}
\end{document}